\documentclass[manuscript,screen]{acmart}
\PassOptionsToPackage{table,xcdraw}{xcolor}
\usepackage{listings}
\usepackage{tikz}
\newcommand{\blackcircle}[1]{%
  \tikz[baseline=(char.base)]{
    \node[shape=circle,fill=black,inner sep=0.5pt,text=white] (char) {#1};
  }%
}

\usepackage{tcolorbox}
\usepackage{tabularray}
\usepackage[table,xcdraw]{xcolor}
\usepackage{multirow}

\usepackage{amssymb}
\usepackage{float}

\AtBeginDocument{%
  }

\setcopyright{acmlicensed}
\copyrightyear{2018}
\acmYear{2018}
\acmDOI{XXXXXXX.XXXXXXX}

\acmConference[Conference acronym 'XX]{Make sure to enter the correct
  conference title from your rights confirmation emai}{June 03--05,
  2018}{Woodstock, NY}
\acmISBN{978-1-4503-XXXX-X/18/06}

\begin{document}

\title{A Comprehensive Study on Dark Patterns}

\author{Meng Li}
\affiliation{%
  \institution{Shenzhen Technology University}
  \city{Shenzhen}
  \country{China}
}

\author{Xiang Wang}
\affiliation{%
  \institution{Shenzhen Technology University}
  \city{Shenzhen}
  \country{China}
}

\author{Liming Nie}
\authornote{Corresponding authors.}
\affiliation{%
  \institution{Shenzhen Technology University}
  \city{Shenzhen}
  \country{China}
}

\author{Chenglin Li}
\affiliation{%
  \institution{Zhejiang Sci-Tech University}
  \city{Hangzhou}
  \country{China}
}

\author{Yang Liu}
\affiliation{%
  \institution{Nanyang Technological University}
  \city{Singapore}
  \country{Singapore}
}
\author{Yangyang Zhao}
\affiliation{%
  \institution{Zhejiang Sci-Tech University}
  \city{Hangzhou}
  \country{China}
}
\author{Lei Xue}
\affiliation{%
  \institution{Sun Yat-sen University}
  \city{shenzhen}
  \country{China}
}
\author{Kabir Sulaiman SAID}
\affiliation{%
  \institution{Aliko Dangote University of Science and Technology}
  \city{Kano}
  \country{Nigeria}
}
\renewcommand{\shortauthors}{Trovato et al.}

\begin{abstract}
As digital interfaces become increasingly prevalent, certain manipulative design elements have emerged that may harm user interests, raising associated ethical concerns and bringing dark patterns into focus as a significant research topic. Manipulative design strategies are widely used in user interfaces (UI) primarily to guide user behavior in ways that favor service providers, often at the cost of the users themselves. This paper addresses three main challenges in dark pattern research: inconsistencies and incompleteness in classification, limitations of detection tools, and insufficient comprehensiveness in existing datasets. In this study, we propose a comprehensive analytical framework—the Dark Pattern Analysis Framework (DPAF). Using this framework, we developed a taxonomy comprising 68 types of dark patterns, each annotated in detail to illustrate its impact on users, potential scenarios, and real-world examples, validated through industry surveys. Furthermore, we evaluated the effectiveness of current detection tools and assessed the completeness of available datasets. Our findings indicate that, among the 8 detection tools studied, only 31 types of dark patterns are identifiable, resulting in a coverage rate of just 45.5\%. Similarly, our analysis of four datasets, encompassing 5,561 instances, reveals coverage of only 30 types of dark patterns, with an overall coverage rate of 44\%. Based on the available datasets, we standardized classifications and merged datasets  to form a unified image dataset and a unified text dataset. These results highlight significant room for improvement in the field of dark pattern detection. This research not only deepens our understanding of dark pattern classification and detection tools but also offers valuable insights for future research and practice in this domain.
\end{abstract}



\received{20 February 2007}
\received[revised]{12 March 2009}
\received[accepted]{5 June 2009}

\maketitle

\section{Introduction}
The emergence of digital interfaces has fundamentally changed the way we interact with technology, giving rise to a series of ethical issues. Among these, "dark patterns" have emerged as a significant focus\cite{Brignull2011,gray2018dark,Gray2023,SanchezChamorro2023,Long2023}. These manipulative design strategies permeate user interfaces (UI) with the aim of guiding user behavior to achieve outcomes favorable to service providers, often at the expense of users\cite{Kollmer2023,Mansur2023a,Porcelli2023,Ramirez2024,Dubiel2024}. Given their widespread presence and impact on user choices and privacy, dark patterns have attracted considerable regulatory attention\cite{bosch2016tales,Dickinson2023,kowalczyk2023understanding,Mildner2023,Kennedy2024}. However, attempts to address this issue have been hindered by a lack of comprehensive understanding and classification frameworks, highlighting the importance of establishing a robust taxonomy and analytical framework.

Dark patterns have increasingly become a focal point of academic research due to their ethical and practical implications across various digital platforms\cite{gray2023towards}. Harry Brignull's pioneering work in 2010 laid the groundwork for this field by providing an initial classification system\cite{brignull2010types}. Current research primarily focuses on two areas: classification and detection. Since its inception, the classification domain has seen significant development, evolving from simple taxonomies to more detailed categories that consider various digital platforms and the cognitive biases exploited\cite{Gray2023c,Kollmer2023a}. Nevertheless, efforts continue to develop a universally accepted classification system that encompasses newly identified dark patterns\cite{di2020ui,gray2018dark,Mathur2019,Gray2024}. Similarly, detection methods are also evolving\cite{chen2023unveiling,Kirkman2023,mansur2023aidui,Sazid2023}. Initial approaches heavily relied on manual reviews, which, while accurate, were impractical due to their time-consuming nature\cite{hidaka2023linguistic}. Subsequent semi-automated methods employed clustering techniques to simulate user behavior\cite{Mathur2019}. Recent work has explored machine learning methods tailored to specific contexts, such as cookie banners, but these require labeled training data\cite{mansur2023aidui}.

However, current research faces several limitations. First, existing classification standards are inconsistent and incomplete. Most studies are confined to specific types or contexts, lacking a comprehensive taxonomy. Furthermore, existing classifications often overlook the impact on users and the scenarios in which dark patterns may arise. Second, there are limitations in detection tools. Current tools often suffer from incomplete coverage or low accuracy. No studies have yet analyzed the full capabilities of these tools within a complete taxonomy. Third, concerns regarding data integrity persist. Previous research has often relied on limited datasets that may not be representative and have notably lacked a comprehensive analysis of their coverage and utility.

In this paper, we introduce the Dark Pattern Analysis Framework (DPAF), a novel two-phase approach designed to address existing gaps in the field. Initially, the framework employs a systematic literature review to assess the current state of dark pattern classification. Subsequently, we embark on constructing and annotating a taxonomy. Specifically, we provide a relatively comprehensive classification of dark patterns that not only integrates existing classifications but also adds types not included in previous taxonomies. Each type is annotated with its impact on users, potential scenarios, and illustrative examples. Following this, we evaluate the capabilities of existing dark pattern detection tools and examine the extent to which current datasets cover the dark pattern types identified in our newly constructed taxonomy, along with some merging and adjustments of existing datasets. Through this approach, we aim to answer three research questions (RQs):

\begin{itemize}
    \item \textbf{RQ1}:Are current dark pattern taxonomies comprehensive? Our systematic review highlights the incompleteness of existing dark pattern taxonomies. To fill this gap, we have created the most detailed dark pattern taxonomy to date, encompassing 68 unique types, validated through industry surveys. We also annotated each type based on its impact on users and potential scenarios, enhancing the taxonomy's applicability, and provided visual and textual examples covering 38 types of dark patterns to facilitate understanding.
    \item \textbf{RQ2}:What are the capabilities and limitations of current dark pattern detection tools? Among the 68 dark pattern types, the selected eight tools can only identify 31 types, resulting in a coverage rate of just 45.5\%. Moreover, these tools fail to recognize the remaining 37 types, accounting for 54.5\% of the total.
    \item \textbf{RQ3}:What is the current status of available data supporting automated dark pattern recognition? Four available datasets contain 5,561 instances but only cover 30 out of the 68 dark patterns, yielding a coverage rate of 44\%. The remaining 38 types lack instances, limiting the capabilities of current detection tools. Additionally, we adjusted the classification of some datasets based on our taxonomy, resulting in a new standard image dataset and text dataset.
\end{itemize}

By answering these questions, this paper deepens our understanding of dark pattern classification and detection tools, and provides valuable insights for future research and practice in this field. For further details, please visit our online website. Our contributions are as follows:
\begin{itemize}
    \item We conducted the first systematic literature review specifically on dark patterns, established a relatively comprehensive dark pattern classification method, and provided examples for some dark pattern types to promote the understanding of dark pattern types.
    \item We summarized the existing dark pattern automatic detection tools and conducted in-depth analysis from three dimensions: the types of dark patterns that can be detected, the effectiveness indicators of detecting a certain type, and the differences in input forms, indicating the capabilities and limitations of current tools.
    \item We sorted out the existing dark pattern detection datasets and analyzed them from three aspects: the scale of the dataset, the instance types of the dataset, and the types of dark patterns covered by the dataset, revealing the usability of the existing datasets and pointing out the direction for the construction of future dark pattern datasets.
    \item Combined with the analysis of the datasets, we adjusted and merged the datasets based on the taxonomy in this paper, forming a standard image dataset and a standard text dataset, which are convenient for subsequent research.
\end{itemize}

It is worth noting that this paper is an extended version of the 2024 ACM FSE paper\cite{nie2024}. In this conference paper, we constructed a new dark pattern taxonomy and preliminarily evaluated existing tools and datasets based on the taxonomy. Although this conference paper describes our process, we also found some shortcomings, such as the lack of display of some dark pattern types, the performance evaluation of detection tools only considering the detected types, etc., and received feedback from reviewers. We made improvements in the following aspects:

First, considering the understanding and update of the taxonomy types, we collected 37 new related papers and obtained 4 new dark pattern types. At the same time, we added the example display section of the dark pattern types to promote the intuitive understanding of the dark pattern types. Secondly, we added three new tools and conducted additional analysis, and conducted further in-depth analysis on the indicator data and input types of the tools to better reveal the capabilities and limitations of the current detection tools. Finally, we carefully reviewed and checked the available datasets, supplemented the annotations of the previously unobtained datasets, adjusted and merged the available datasets based on the taxonomy types, unified the classification labels of the current datasets, and facilitated the use of subsequent research.
\section{Background and Motivation}

\subsection{Dark Patterns}
\begin{figure}[h]
\centering
\begin{minipage}[t]{0.4\textwidth}
            \centering
			\includegraphics[width=1.6in]{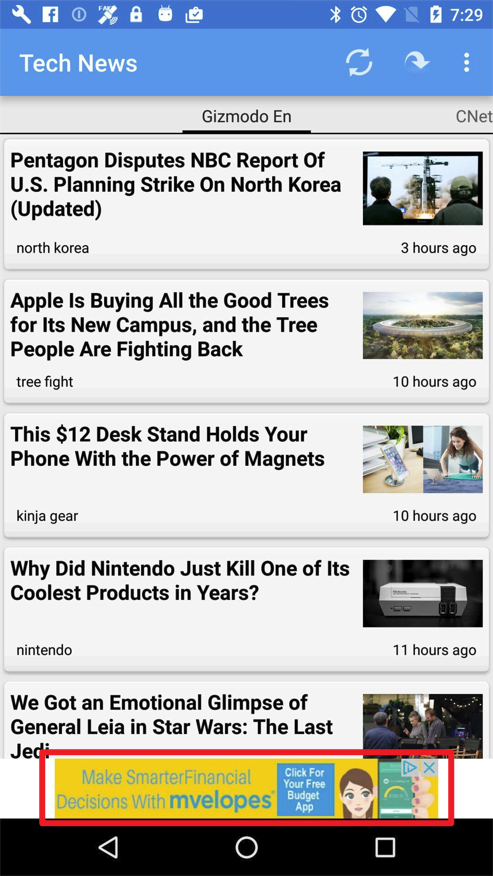}
                \centerline{\footnotesize{Small close button}}
                \vspace{-5pt}
		\end{minipage}
		\begin{minipage}[t]{0.4\textwidth}
			\centering
			\includegraphics[width=1.6in]{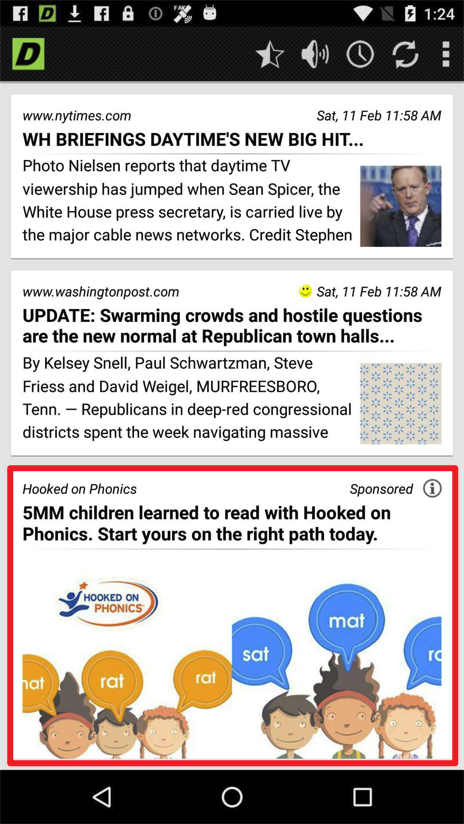}
            \centerline{\footnotesize{Disguised ad}}
		\end{minipage}
\vspace{-5pt}
\caption{Some examples of Dark Pattern}
\vspace{-5pt}
\label{fig:examples}
\end{figure}

In today's society, user interface (UI) and user experience (UX) design have become deeply integrated into our daily lives. Whether we are shopping, socializing, or working, we constantly interact with various applications and online platforms. The design of these platforms directly influences our decisions, emotions, and behaviors. A well-designed interface can simplify tasks, enhance user experience, and even encourage consumers to make more informed choices. However, with technological advancements and increasing commercial competition, a design strategy known as "dark patterns" has emerged.
The term "dark patterns" was introduced in academic literature\cite{Boring2014,brignull2015dark} and refers to manipulative or deceptive elements in UX design. These are intentionally crafted designs aimed at inducing users to take actions they might not willingly choose if they were fully informed or aware of other options\cite{Mathur2019}. This strategy exploits individuals' cognitive biases and habitual behaviors to achieve specific commercial or other goals, such as tricking users into clicking on ads, agreeing to subscriptions unknowingly, or sharing personal data without complete understanding. Dark patterns manifest in various forms, such as misleading buttons, hidden fees during checkout, complicated unsubscribe processes, and ads disguised as genuine content. For instance, Figure\ref{fig:examples} shows two dark patterns. The left picture shows ads that often appear at the bottom of mobile apps. The close button is very small (marked in a red box). Users need to click repeatedly to close the ads, and may even accidentally click on the ad page, which affects the user's autonomy and wastes time. The right picture shows disguised ads (marked in a red box) mixed in with the normal content provided by the app, trying to trick users into clicking. These design practices can erode trust in products or services and may even lead to legal liabilities, posing potential adverse impacts on users, businesses, and the broader online ecosystem\cite{fansher2018darkpatterns,zagal2013dark,Long2023}. Fundamentally, dark patterns are considered unethical design strategies because they manipulate and deceive users, forcing them to make choices that benefit online services, often at the expense of the users' best interests.

\subsection{Motivation}
The increasing prevalence of dark patterns in user interfaces has garnered significant attention. These deceptive designs undermine user trust, prompting decisions that are not in the users' best interests, potentially reducing user engagement and retention. Beyond manipulation, dark patterns can inadvertently lead users to make unintended purchases or share sensitive information, posing legal challenges for companies and infringing on user rights. Additionally, by creating a sense of urgency, these patterns increase psychological stress, potentially evolving into broader mental health issues\cite{Hilton2023DarkPA}. From a data security perspective, dark patterns heighten the risk of data breaches when users are deceived into providing excessive information. While dark patterns may improve short-term metrics, their extensive and adverse long-term impacts on individuals and society warrant serious consideration.

The widespread use and potential negative impacts of dark patterns underscore the urgent need for a better understanding and identification of these deceptive practices. This is crucial for preventing their proliferation and enhancing the overall quality of software. However, addressing dark patterns presents numerous challenges in terms of technology, concepts, and practices. The main obstacles include:

\begin{itemize}
\item \textit{Inconsistencies and Incompleteness in Classification}:
Despite significant research efforts to create classification frameworks for dark patterns, many of these efforts are limited to certain types or contexts, lacking a unified and standardized classification. A comprehensive taxonomy that encompasses various sources of dark patterns has yet to be established. This lack of standardization hinders the investigation, identification, and addressing of these manipulative design practices.
\item \textit{Limitations of Detection Tools}:
Given the potential harm caused by dark patterns, the creation and implementation of automated detection tools are crucial. These tools can act as deterrents, encouraging designers and developers to adhere to ethical and user-centric design principles. However, the effectiveness of current tools varies, with many struggling to achieve comprehensive coverage or maintain accuracy. A critical evaluation of these tools' capabilities and limitations is essential for advancing research and improvements in this field. Despite the clear need for such evaluations, dedicated studies on these aspects are notably lacking.
\item \textit{Inadequacies in Data Comprehensiveness}:
The effectiveness of dark pattern detection tools largely depends on the quality and size of the datasets they rely on. Deficiencies or biases in these datasets can reduce the tools' effectiveness. Many studies utilize limited datasets that may not fully represent broader scenarios and lack comprehensive analysis of their content and applicability. Therefore, evaluating current datasets with a focus on their diversity and comprehensiveness is crucial.
\end{itemize}

Given these existing challenges, a comprehensive exploration of dark patterns becomes imperative. Driven by this urgency, our research aims to bridge the existing knowledge gaps, provide a robust unified classification standard, and conduct a detailed review of current dark pattern detection tools and available data. Through our efforts, we seek to advance academic discourse and practical interventions related to dark patterns.

\section{Methodology}

In this section, we first outline the three research questions and the intentions behind them. Next, we introduce the overall framework of the corresponding empirical study. We then describe each component and step within the proposed framework in detail, with a particular focus on building a taxonomy of dark patterns, annotating tool capabilities and limitations, and marking data availability, as well as the process of creating a standard dataset.
\subsection{Research Questions(RQs)}

\begin{itemize}
    \item RQ1: Are current dark pattern taxonomies comprehensive?To effectively understand, detect, and mitigate dark patterns, a comprehensive and widely accepted taxonomy is essential. A clear and complete taxonomy enables researchers and practitioners to better understand the diversity and complexity of dark patterns. This, in turn, supports the development of relevant policies and regulations and enhances the effectiveness of detection tools. Therefore, constructing the most comprehensive and industry-recognized taxonomy of dark patterns is critical.
    \item RQ2: What are the capabilities and limitations of current dark pattern detection tools?Detecting dark patterns is time-consuming and requires expert knowledge. Although several automated detection tools have been developed, their capabilities and limitations remain insufficiently studied or understood. A thorough investigation into these aspects is crucial for optimizing these tools, improving their accuracy and reliability, and expanding their application scope.
    \item RQ3: What is the current status of available data supporting automated dark pattern recognition?A significant barrier to research and solutions related to dark patterns is the lack of comprehensive and reliable datasets. However, the current state of data availability has not been adequately explored. Therefore, an in-depth analysis of the availability of existing data and its actual coverage within a comprehensive dark pattern taxonomy is necessary and urgent. Additionally, by adjusting and merging existing datasets based on the taxonomy presented in this paper, we aim to establish a solid foundation for future research.
\end{itemize}

\begin{figure}[h]
\centering
\includegraphics[width=1\linewidth]{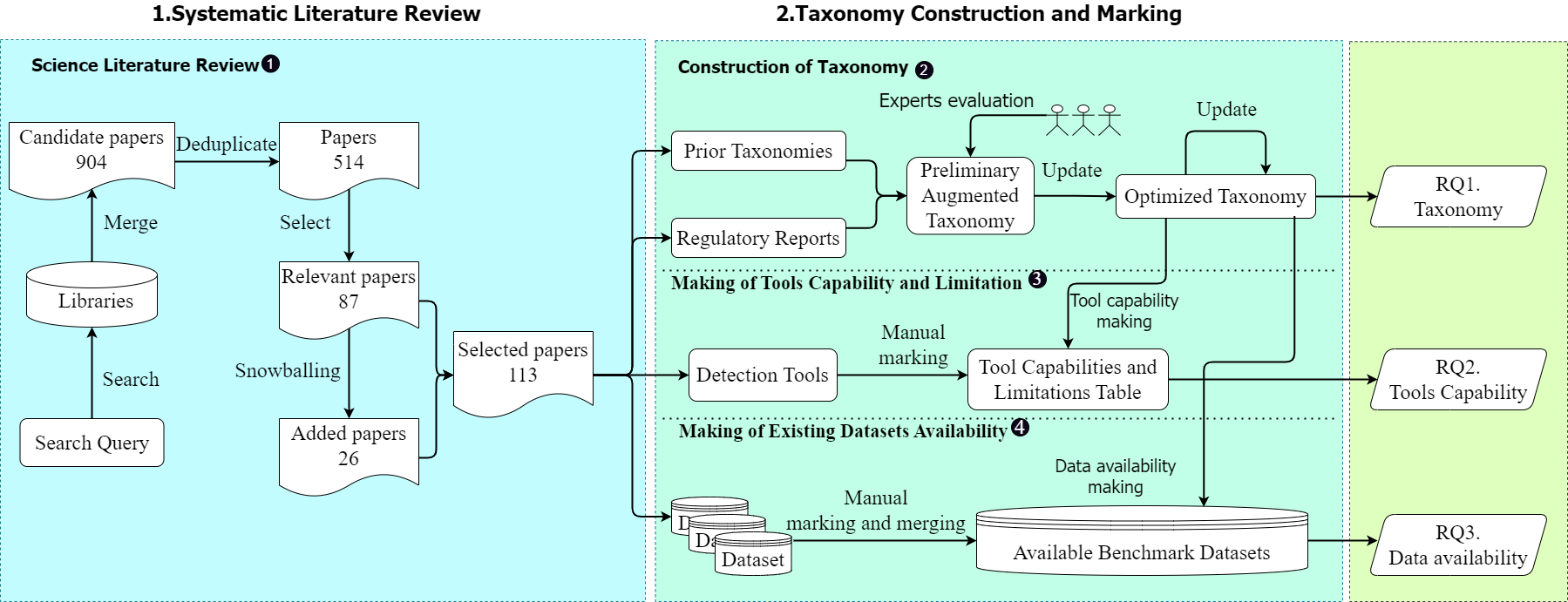}
\vspace{-10pt}
\caption{The framework of DPAF}
\vspace{-15pt}
\label{fig:framwork}
\end{figure}

\subsection{Framework}

To address the three research questions identified above, we propose a framework named the Dark Pattern Analysis Framework (DPAF), based on the taxonomy construction methods outlined by Ladisa et al.\cite{ladisa2023sok}. Our framework encompasses building a taxonomy of dark patterns, annotating the capabilities and limitations of existing detection tools within this taxonomy, marking data availability, and establishing an update mechanism. As illustrated in Figure ~\ref{fig:framwork}, DPAF consists of two stages: a systematic literature review and taxonomy construction and annotation. Each stage and its specific steps are discussed in detail below.

\subsection{Scientific Literature Review}

This section describes the systematic review of academic literature related to dark pattern research \blackcircle{1}. We adopted a structured approach for a literature review, covering various stages\cite{kitchenham2009systematic,nie2023systematic}.

\subsubsection{Design of Search Query}
Our systematic review on dark patterns began by designing an effective search query. To ensure the reliability of the query, we conducted an exploratory search on Google Scholar using the keyword “dark pattern.” This preliminary step revealed four relevant studies\cite{Brignull2011,Kitkowska2023,mansur2023aidui,Mathur2019}, providing us with an initial understanding of the topic and helping us identify relevant keywords. We then analyzed these keywords to construct a search query that would retrieve the most pertinent papers on dark patterns. The final search query was as follows:

\begin{lstlisting}
(Dark patterns OR dark nudging OR Digital sludging OR Manipulation techniques OR Deception techniques OR deceptive button OR dark pattern recognition OR dark pattern classification AND GUI OR UI OR user interface OR graphical user interface)
\end{lstlisting}

\subsubsection{Literature Search}
With a carefully designed search query, we conducted a comprehensive literature search across four renowned databases: IEEE Xplore, Scopus, Google Scholar, and the ACM Digital Library. We restricted our search to the title, abstract, and author keywords, focusing only on English publications, including journal articles, conference proceedings, and book chapters.

\subsubsection{Merge Candidate Literature}
Following the systematic search, we recorded the number of candidate papers retrieved from each database: Google Scholar (509), IEEE Xplore (147), Scopus (129), and ACM Digital Library (119). In total, we initially identified 904 candidate papers.

\subsubsection{Deduplication}
To ensure the completeness of our review, we meticulously removed duplicate entries from the candidate literature pool. This process yielded a refined set of 514 papers, which then proceeded to the next stage of paper selection based on inclusion criteria.

\subsubsection{Paper Selection}
The paper selection process was conducted in two stages. In the initial stage, we conducted a preliminary screening by closely examining the metadata of each paper, paying particular attention to publication type, title, and abstract. Papers that were clearly unrelated to dark patterns in user interfaces were excluded. This initial screening reduced the candidate pool to 183 papers, which advanced to the second stage of selection.
In the second stage, we obtained the full text of each remaining paper and conducted a comprehensive review of its content. Specifically, we examined the introduction, methods, and results sections to confirm whether each paper presented research related to dark patterns in user interfaces. Additionally, all sections of the studies had to be written in English to be considered. Papers that did not meet these criteria were excluded. To enhance the accuracy and consistency of our review, we employed a parallel review strategy where each paper was independently reviewed by at least two authors. Inclusion in the final review required consensus among the authors regarding the paper’s relevance. At the end of this rigorous screening process, we identified 87 relevant papers.

\subsubsection{Snowballing}
To further enhance the completeness of our selection, we employed a bidirectional snowballing strategy, tracing both forward and backward references from the initial 87 papers deemed relevant. This method applied the established screening criteria consistently across all rounds of snowballing. We included only documents directly cited by or citing our selected literature, limiting the inclusion to a maximum of two citation levels to maintain focus and manageability. This iterative process resulted in the addition of 26 new papers. Consequently, we finalized a selection of 113 relevant papers, forming the foundation of our comprehensive study on dark patterns.

\begin{figure}[h]
\centering
\includegraphics[width=0.8\linewidth]{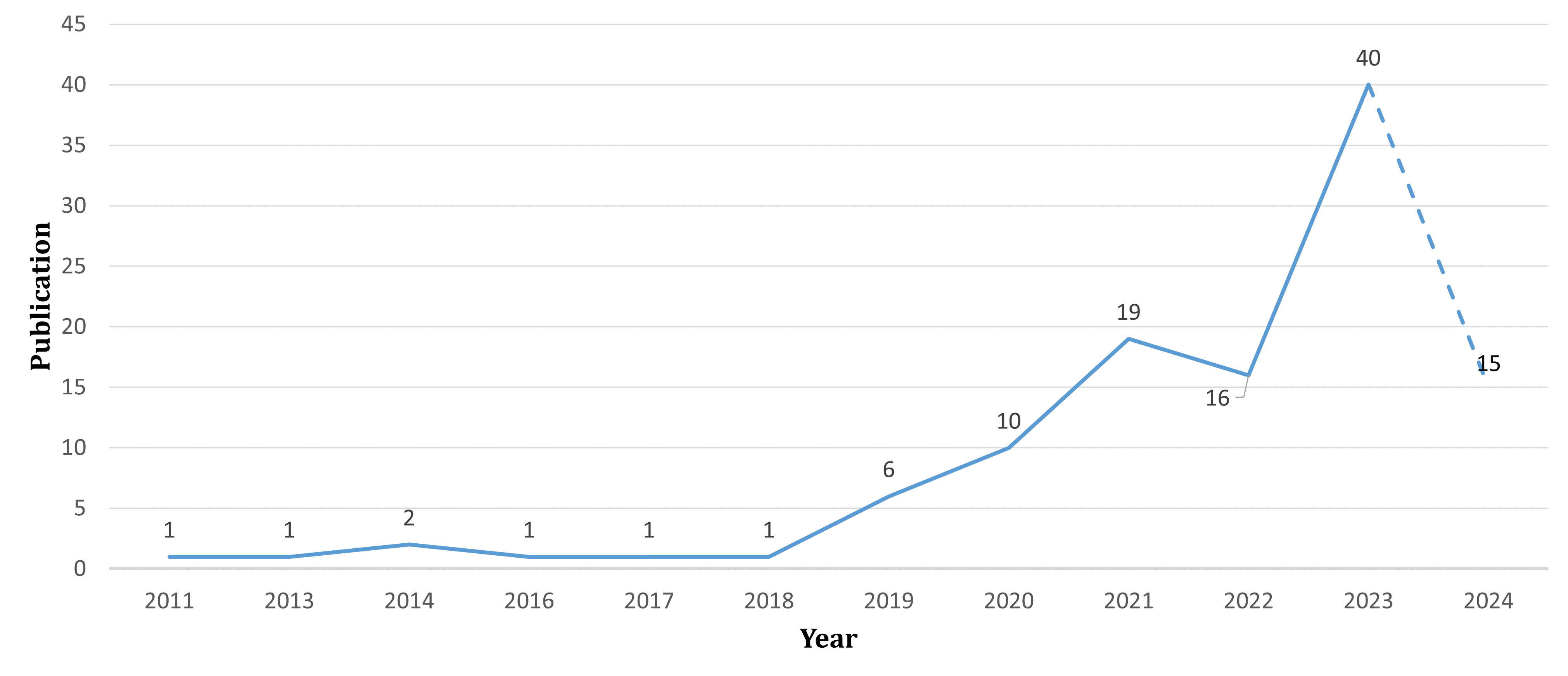}
\vspace{-10pt}
\caption{Annual Publication Trend of Dark Patterns-Related Research Papers}
\vspace{-15pt}
\label{fig:publicationtrend}
\end{figure}

Figure \ref{fig:publicationtrend} illustrates the publication trends by year based on these 113 selected papers, showing an overall upward trend (with only the first four months of publications considered for 2024). These publications were meticulously reviewed to extract details related to the prevalence, identification, and classification of dark patterns, providing the data needed to address our research questions.

\subsection{Construction of Taxonomy}
This section describes the methodology for developing an optimized taxonomy\blackcircle{2}, as shown in Figure~\ref{fig:framwork}, representing an industry-recognized augmented version. To achieve this, the taxonomy construction process is divided into three stages: first, constructing an initial augmented taxonomy as an enhanced version of existing taxonomies; second, conducting an industry survey to evaluate the rationality, completeness, and usefulness of this preliminary taxonomy; and finally, refining the taxonomy based on expert feedback to produce the final optimized version.

\subsubsection{Building an Preliminary Augmented Taxonomy}

The main objective of this step is to build an enhanced version of the taxonomy based on previous work on dark patterns. Previous research includes prior taxonomies\cite{bosch2016tales,brignull2015dark,gray2018dark,gray2023towards,luguri2021shining,Mathur2019,Gray2024,Lacey2023,Saville2024}, reports\cite{CMA,EDPB,FTC,EUCommission,OECD} , as well as some dark patterns mentioned in papers but not included in prior taxonomies\cite{bosch2016tales,chen2023unveiling,DiGeronimo2020,gray2020kind,gunawan2021comparative,lacey2019cuteness,mansur2023aidui,zagal2013dark,Stavrakakis2021,Singh2022}.
Additionally, we annotated each dark pattern in the preliminary augmented taxonomy regarding its potential impact on users, application scenarios, and available type examples. The goal is to gain a more comprehensive understanding of the impact of dark patterns on users, as well as the scope and specific type instances of their application across various contexts. To achieve these objectives, we followed a multi-step approach divided into four main phases.

\noindent\textbf{Aggregation of Existing Taxonomies. }
First, from the 113 relevant articles, we identified 18 key taxonomies that provided substantial foundations for our work. We manually extracted existing dark pattern types and their classification information from these prior taxonomies.

Second, using Gray et al.'s taxonomy\cite{gray2023towards} as a baseline, we integrated similar dark patterns to avoid redundancy. This integration was divided into two categories: textual similarity and conceptual similarity. On one hand, textual similarity refers to those dark patterns that either have identical definitions, point to the same instances, or use synonyms. For these types of patterns, we only needed to record their source papers, descriptions, and type classifications from the original taxonomy versions. For those with category inconsistencies, three authors discussed and determined the final category based on the existing classifications and the majority rule. On the other hand, conceptual similarity indicates patterns that differ in terminology but are essentially similar. Since we could not know in advance which dark patterns were conceptually similar, we used manual verification to check all patterns beyond textual similarity. Specifically, two authors independently read each pattern’s text description, relevant examples, and classification information, providing their own conclusions on whether they belonged to the same dark pattern or should be classified separately. After both authors completed all annotations, we identified patterns with annotation discrepancies. For these annotations, a third author was invited to make an independent judgment, followed by a final consensus. The result of this stage was an initial version of the most comprehensive taxonomy covering all known dark patterns.

Conceptual similarity is applied for assessing more complex cases of similar dark patterns. It primarily involves the following steps: a) Judgment based on similarity of descriptions: For definitions of dark patterns, although they may be described in different words, they may convey the same or similar meanings. These types of dark patterns can be merged into the same category. For all detection tools, the corpus is an important basis for detecting dark patterns. Therefore, this step aligns with a classification method based on the foundations of detection. b) Judgment based on the similarity of typical examples: Most existing detection tools use visual cues as a key reference for detecting dark patterns. Therefore, we focused on whether typical examples of two dark patterns are similar in their presentation, for example, in their graphical interface and descriptions. c) Judgment based on potential impact on users: To stay consistent with real-world discussions on dark patterns, we considered the different types of impacts that dark patterns may have on users as an additional evaluation step. The authors carefully reviewed five guidelines proposed by regulatory bodies or organizations. These articles provided summaries of dark patterns, and some of their insights were incorporated into our dark pattern taxonomy construction.

Furthermore, in previous dark pattern classification methods, nearly all existing detection tools use the smallest unit of dark pattern types as the detection unit. Since they are widely discussed, they were retained as representative types in the taxonomy construction process, meaning they were typically not merged with other dark pattern types. Given the importance of classification for detection tools, our website provides specific examples of these consolidations to illustrate the process of merging dark pattern types more effectively.

\noindent\textbf{Add Missing Dark Patterns.}
In building upon the initial taxonomy of dark patterns, which aggregated findings from prior literature, our objective is to construct a more comprehensive taxonomy to date. Despite the extensive categorization already provided in existing taxonomies, such as the seminal work by Gray et al.\cite{gray2023towards}, we identified coverage gaps that may have resulted from divergent focal points in previous research. To rectify this, we undertook a deep reading of all relevant literature, comprising 113 papers, in order to broaden the scope of our taxonomy. Specifically. we have supplemented the latest taxonomy from\cite{gray2023towards} by incorporating additional 10 dark patterns, thereby establishing a preliminary taxonomy for the upcoming survey phase.

\noindent\textbf{Assessment of User Impact.}
Although much of the prior research has primarily focused on the identification and description of dark patterns, there has been a comparatively limited exploration of the end impact these patterns impose on consumers. In an attempt to offer a comprehensive understanding of the multi-dimensional ramifications that all variants of dark patterns could potentially exert on users, we engage in a systematic annotation of each identified type of dark pattern with its associated harms. For the purposes of this study, we employ the foundational framework proposed by the Organization for Economic Cooperation and Development (OECD) to categorize the harms of dark patterns\cite{OECD}. This framework identifies dark pattern harms from six unique vantage points, which, for ease of reference and clarity, we herein label as H1 to H6: 

\begin{itemize}
\item \textit{Harm to User Autonomy (H1)}: This category suggests that certain dark patterns compromise user autonomy by forcing consumers to make choices they would not otherwise make, limiting the available options, and obfuscating the decision-making process.

\item \textit{Personal User Detriment}: This domain encompasses three separate, yet interconnected, harms: \textit{Financial Loss} (H2), \textit{Privacy Harms} (H3), and \textit{Psychological Detriment and Time Loss} (H4). H2 addresses dark patterns that manipulate consumers into making unnecessary purchases or overspending. H3 emphasizes how dark patterns could induce users into unintentionally disseminating excessive personal information, thereby elevating their risk exposure. H4 reflects on dark patterns that place emotional and cognitive strain on users, exploiting their vulnerabilities and potentially leading to time wastage.

\item \textit{Structural User Detriment}: This facet includes two distinct, yet interrelated,  harms: \textit{Weaker or Distorted Competition} (H5) and \textit{Reduced Consumer Trust and Engagement} (H6). H5 means that dark patterns can distort market competition by preventing or dissuading consumers from shopping and comparing offers. H6 posits that dark patterns may erode consumer trust and engagement by deceiving users into oversharing information or overspending, particularly if such practices are subsequently uncovered.

\end{itemize}

In alignment with the aforesaid definitions, we performed a meticulous assessment of each identified dark pattern to gauge its potential impact on users. This process consisted of two pivotal phases: First, we unambiguously included those dark patterns that were explicitly cited by the OECD as emblematic instances of particular kinds of harm. These were directly incorporated into our analytical framework. Subsequently, for any unaccounted for dark patterns that were not explicitly delineated, our interdisciplinary team of three authors undertook rigorous collective discussions. These discussions aimed to systematically assign potential harms to users, guided by the definitions and corresponding examples established by the OECD\cite{OECD}.

\noindent\textbf{Common Scenarios.}
To gain a more comprehensive understanding of the possible impacts of dark patterns on users, particularly in different application scenarios, it is imperative to annotate the scenarios in which these patterns manifest. Such annotation serves a dual purpose: it not only identifies the broad applications of dark patterns but also delves deeper into their potential impacts within diverse user groups and environments. Towards this aim, we have deployed a two-fold methodological approach for systematically annotating various dark pattern scenarios. 

Initially, we amalgamated dark pattern scenarios discussed in previous research\cite{hidaka2023linguistic,kowalczyk2023understanding} and categorized them into four distinct classes: Social Platforms (S), E-commerce (E), Entertainment Software (ES), and General or Universal (All). These classifications serve as a robust foundation for the subsequent annotation of dark patterns. In the second step, we embark on an intricate and detailed annotation process. Initially, we extract pertinent data regarding the potential impact of dark patterns across various scenarios from existing scholarly publications. To elaborate, suppose a particular paper concentrates on categorizing types of dark patterns prevalent on social platforms\cite{schaffner2022understanding}. During our data acquisition phase, we would not only incorporate all types of dark patterns detailed in that paper into our taxonomy but also label them under the dimension of Social Platform to denote the scope of their impact. This approach might result in a specific type of dark pattern being discussed in multiple articles centered around different application scenarios, thus acquiring multiple tags. This is, in fact, an expected outcome. Our taxonomy specifically identifies three key application scenarios: Social Platforms, E-commerce, and Entertainment Software. Therefore, should a dark pattern type be prevalent across these three scenarios, we assign it the designation All. 

At this point, we can complete the annotation of all dark pattern type impact scenarios, thereby furnishing a nuanced and comprehensive framework that enhances our understanding of how these patterns operate across various contexts.

\noindent\textbf{ Instance Collection.}
After annotating the impact and application scenarios of each dark pattern type, we further visualized these types by collecting and displaying specific instances. This step provides a tangible understanding of the scope and influence of dark patterns.

First, we defined the scope for data collection. Through literature review, we observed that instances of dark patterns primarily appear as images and text. Therefore, our collection focused on images and text illustrating dark patterns. Images include screenshots of mobile apps or web pages displaying dark patterns, while text refers to English phrases on these platforms that capture the essence of a dark pattern. Additionally, certain studies on dark pattern detection provide relevant datasets. We thoroughly examined these datasets, verifying and collecting instances based on their labels. In our taxonomy, we marked the availability of instances using “I” for image instances, “T” for text instances, and “-” for unavailable instances. The complete list of collected instances is shown in our website, which record each dark pattern type, a description of the type (based on Gray's research\cite{Gray2024}), specific instances, and their explanations.

After successfully completing these key steps, we set an abbreviation for each type in the taxonomy, which is mainly composed of the first two letters of \textit{“Categories”} and the first two letters of \textit{“Dark Pattern Types”} or \textit{"Sub-Categories"}, to facilitate the subsequent use of the type and its positioning in the taxonomy.Ultimately, we developed a comprehensive initial taxonomy of existing dark patterns. This taxonomy encompasses all dark patterns referenced in current academic research and regulatory reports and expands on three additional dimensions: severity, impact scenarios, and collected instances. These elements enhance our understanding of the potential impact of dark patterns on users, the various contexts they affect, and corresponding real-world examples.

\subsubsection{Conducting the Expert Survey}

The primary aim of this online survey is to gather feedback from industry professionals, with a particular focus on designers and UI-related developers. The survey addresses multiple focal points: a).Assess the rationality, completeness, understandability, and usefulness of the dark pattern taxonomy conducted by our work. b). Assess the effectiveness of the annotations in terms of the degree of impact on users and common scenarios. c).Assess the efficacy of existing tools that detect dark patterns. d).Assess the availability of existing dark pattern datasets. e).Collect information from practitioners on their awareness and strategies for identifying and mitigating dark patterns.

The entire survey construction process consists of the following three parts: The first is the design of the questionnaire. The questionnaire comprises 16 questions. Various aspects of focal points are evaluated using a Likert scale\cite{albaum1997likert}, ranging from 1 (low) to 5 (high). A minority of questions use binary options, such as whether participants have actually read the taxonomy document and background materials provided by us. Another binary question assesses whether participants feel adequately informed to provide valid feedback. These are critical factors for the quality of the survey data. Additionally, the questionnaire also includes demographic questions that cover participants’ age, tenure in their respective industries, and background information. The second is the distribution and scoring method. Snowball sampling techniques are utilized to reach potential participants\cite{goodman1961snowball}. Initial recruitment is targeted at designers and UI-related developers with more than 5 years of industry experience, which is later expanded to encompass general practitioners in related fields. The primary channels of recruitment include social media platforms and industry forums. Participants are required to first read the taxonomy document and background materials provided by us and subsequently rate each addressed question. The third is privacy safeguards. To ensure the confidentiality and anonymity of participants, the survey does not collect personally identifiable information. The data collection process adheres to the principle of data minimization. Responses are securely stored, and only aggregated data is reported. The survey is developed using Wjx.cn, a comprehensive platform for survey design, data collection, and analysis. For the actual setting of the questionnaire question, please refer to our website. The survey started on September 1, 2023, and ended on September 25, 2023. The survey collected feedback from 176 respondents.

\subsubsection{Updating the Taxonomy}
Upon receiving feedback from participants, we will engage in a systematic process to update our existing taxonomy. Initially, we will analyze all the responses collected, meticulously reviewing any contentious points or suggested modifications. Then more interviews with industry experts will be conducted to determine whether the proposed changes should be implemented. Second, any alterations to the taxonomy will be rigorously documented, with annotations explaining the rationale behind each modification. This framework ensures that the taxonomy remains dynamic and evolves in accordance with expert opinions and industry needs. Furthermore, to account for the dynamic nature of dark patterns, especially as user interfaces continue to evolve and new types of dark patterns emerge, it is imperative to incorporate a robust maintenance mechanism within DPAF. This mechanism would involve periodic reviews and updates to the taxonomy, facilitated by automated scraping algorithms and expert consultations. In doing so, DPAF remains an adaptive tool that not only captures the current state of dark patterns, but also anticipates and adapts to future developments.

\subsection{Analysis of Detection Tools}\label{sec:selecttools}

We conducted an in-depth analysis of the 113 articles compiled in Section 3.3, aiming to identify all automated tools adept at detecting dark patterns \blackcircle{3}. To streamline our selection process and ensure a focused analysis, we established the following criteria. Only tools meeting all these criteria were included in our study.

\begin{itemize}
\item  \textbf{\textit{Criterion \#1 (Dark Pattern Recognition)}}: It is imperative for the tools to have a direct bearing on dark pattern identification tasks. Their descriptions should encompass verbiage that resonates with keywords such as "identification," "detection," and "classification."

 \item  \textbf{\textit{Criterion \#2 (Automation)}}: 
 Recognizing that manual identification has inherent drawbacks like increased time consumption and reduced efficiency, our analysis zeroes in exclusively on tools offering automated capabilities.

 \item   \textbf{\textit{Criterion \#3 (Functional Description)}}: Clarity is crucial. As such, each tool must be supported by comprehensive and clear documentation, to ensure a seamless evaluation of its features and functionalities.

 \item   \textbf{\textit{Criterion \#4 (Types of Dark Patterns Detected)}}: The documentation of each tool should elucidate the specific dark patterns that it can detect.
\end{itemize}

After applying these criteria, we identified eight representative tools and conducted a detailed analysis covering aspects such as the number of detectable dark pattern types, performance metrics, and input data types. Specifically:First, we recorded each tool’s publication year and platform based on a comprehensive literature review. This information serves as a key basis for evaluating the currency of each tool, while the impact factor of the journal or conference provides an important reference point for the tool's scientific value. Next, we reviewed each tool’s “method” section to list the primary techniques, data input types, and application platforms, providing an overview of the main techniques used for dark pattern detection, the types of data utilized, and common application platforms. Finally, for the types of dark patterns each tool could detect and the corresponding metric data, we applied a careful manual tagging strategy.Specifically, we reviewed each tool (whether open-source or commercial) through a thorough literature review, documentation analysis, and code inspection. Since many tools are not open-source and we did not conduct experimental tests, we recorded the dark pattern types each tool claims to detect and the corresponding metrics. Throughout this process, we systematically annotated each tool’s capabilities based on the taxonomy we created for dark patterns. The primary metrics we recorded were the four most commonly reported: Accuracy, Precision, Recall, and F1-score. We record these metrics from two main aspects, and the specific data will be presented in Table\ref{tab:tools}:

\textit{Overall Performance Metrics}: Based on our review, we recorded each tool's overall performance metrics, reflecting its general efficacy in detecting dark patterns. These metrics, used to compare the performance of various tools, are shown in the first row under each tool name in Table\ref{tab:tools}.

\textit{Detection Metrics for Different Dark Pattern Types}: We also recorded the detection performance metrics of each tool for different dark pattern types, which serve as the basis for analyzing each tool's effectiveness in identifying specific dark patterns. This information appears in Table\ref{tab:tools} under the section listing the detectable types for each tool. When certain metrics repeat, it often indicates that the tool presents results in "Category" classifications. For these cases, we verified the data set to check the types of dark patterns it includes and annotated accordingly. A "-" in the metric data means no data was available, typically due to missing results or lack of complete code for testing.

Based on the taxonomy integration in Section 3.4, if a tool claims to detect a specific dark pattern that has been merged with another similar pattern in our taxonomy, the newly merged pattern will be marked as detectable. By employing this systematic and meticulous approach, we effectively evaluate and compare the performance of various dark pattern detection tools, providing researchers and developers with valuable insights.

\subsection{Analysis of Datasets}

To determine whether current public datasets are sufficient for dark pattern research, we evaluated their comprehensiveness in terms of dark patterns\blackcircle{4}. To achieve this, we set precise criteria for dataset selection, ensuring alignment with the complexity of dark pattern detection.

Initially, we reviewed our collection of 113 dark pattern-related articles to identify any datasets mentioned or implied to contain dark pattern instances. Our dataset selection adhered to two primary criteria:
\begin{itemize}
 \item  \textbf{ \textit{Criterion \#1 (Presence of Dark Pattern Instances)}}: The dataset should explicitly contain screenshots or videos of web pages or mobile apps with dark patterns. This provides researchers with a solid foundation from which to extract and comprehend the features of dark patterns.
 \item  \textbf{\textit{Criterion \#2 (Open Accessibility)}}: The dataset must be publicly disclosed by its authors and readily available without constraints.
\end{itemize}

After applying these criteria, we identified four datasets relevant to dark pattern detection. We then collected and analyzed information about these datasets. First, we reviewed the associated literature to record each dataset's publication references and release year. Additionally, we documented the types of instances and corresponding platforms within each dataset, such as image or text data and mobile application or web-based content. Finally, we compiled statistics on the number of instances and the range of dark pattern types covered in each dataset. Instance count indicates dataset size, while the number of dark pattern types covered reflects the sample diversity. For the types of dark patterns covered in the dataset, we compared the instances in the dataset one by one based on the taxonomy constructed in this paper, through reading and checking the literature, and obtained the number of dark pattern types contained in the dataset and counted the number of instances of each type covered, which is also the basis for our subsequent analysis of the dataset. Consistent with our previous method of marking tool capabilities, if a covered dark pattern has been merged with another dark pattern, we will mark the merged dark pattern as present in the dataset.

\subsection{Construction of Standard Datasets}

Based on our collection and analysis of existing datasets, we found that different datasets have inconsistent classification standards. In order to unify the classification labels of the datasets for the convenience of subsequent research, we adjusted and merged the available datasets based on the classification method proposed in this article to form a new standard dataset.

First, we used manual inspection to carefully review the duplicate and blank data in the dataset. Two authors checked one by one to ensure the accuracy of the data and removed blank and duplicate data. Secondly, based on our description of the dataset annotation in Section 3.6, we checked and adjusted the labels in the dataset. Next, we merged datasets with the same instance form. For example, for two adjusted image datasets, we put the image instances in the two datasets together and merged the labels of the two datasets to form a standard dataset with more instances, wider type coverage and unified annotations.

In this way, we successfully unified the classification standards of the available datasets. The merged new dataset has more instances and richer types, which is conducive to subsequent research use.

\section{Results and Insights}
This section introduces the dark pattern taxonomy built on our proposed DPAF framework, as well as annotations on tool capabilities, data availability, and dataset adjustments on this taxonomy to address the three research questions.

\subsection{Taxonomy}
~
\begin{table*}[]
\centering
\caption{Final Optimized Taxonomy of Dark Patterns}
\label{tab:Taxonomy}
\resizebox{\textwidth}{!}{%
\begin{tabular}{llllllcc}
\hline
\textbf{Categories}              & \textbf{Sub-Categories}                    & \textbf{Dark Pattern Types}                          & \textbf{Abbreviation} & \textbf{Sources}                                                               & \textbf{Severity}                & \textbf{Common Scenarios$^\S$}    & \textbf{Instance}         \\ \hline
Nagging$^\dagger$                          & -                                          & -                                                    & NG                    & \cite{chen2023unveiling,gray2018dark,luguri2021shining,Gray2024,Lacey2023,Kennedy2024}                                                                              & H3,H4,H5                         & ALL                          & I                         \\ \hline
Obstruction                      & Roach Motel                                & Immortal Accounts$^\dagger$                                    & OB-IA                 & \begin{tabular}[c]{@{}l@{}}\cite{bosch2016tales,brignull2015dark,chen2023unveiling,gray2018dark,luguri2021shining,Mathur2019,Gray2024,Kelly2024}\end{tabular}              & H1,H2,H3,H5                      & ALL                          & I/T                       \\
                                 &                                            & Dead End$^\dagger$                                              & OB-DE                 & \cite{EDPB,Gray2024}                                                                                            & H1,H2,H4                         & S                            & -                         \\
                                 &                                            & Forced Grace Period*                                  & OB-FG                 & \cite{Mildner2023}                                                                                               & H1,H4                            & All                          & I                         \\
                                 & Creating Barriers                          & Price Comparison Prevention$^\dagger$                          & OB-PC                 & \begin{tabular}[c]{@{}l@{}}\cite{brignull2015dark,chen2023unveiling,gray2018dark,luguri2021shining,Gray2024,Lacey2023,Saville2024,Stavrakakis2021}\end{tabular}                            & H1,H2,H5,H6                      & E                            & I                         \\
                                 &                                            & Intermediate Currency$^\dagger$                                & OB-IC                 & \cite{chen2023unveiling,gray2018dark,luguri2021shining,Gray2024,Lacey2023}                                                                                & H1,H2,H5                         & S/ES                         & I                         \\
                                 & Adding Steps                               & Privacy Maze$^\dagger$                                         & OB-PM                 & \cite{EDPB,Gray2024}                                                                                            & H1,H3                            & S                            & -                         \\
                                 &                                            &  Labyrinthine Navigation*      & OB-LN                 & \cite{Mildner2023}                                                                                              & H1                               & S                            & -                         \\ \hline
 Sneaking &  Bait And Switch    &  Disguised Ad                 & SN-DA                 &  \begin{tabular}[c]{@{}l@{}}\cite{brignull2015dark,chen2023unveiling,gray2018dark,luguri2021shining,Gray2024,Lacey2023}\\ \cite{Lu2024,Saville2024,Stavrakakis2021}\end{tabular} &  H4,H6    &  ALL  &  I \\
                                 &                                            &  Only Initial Payouts*         & SN-OI                 & \cite{Singh2022}                                                                                               & H2,H5,H6                         & E                            & -                         \\
                                 &                                            &  Cannot Redeem*                & SN-CR                 & \cite{Singh2022}                                                                                               & H2,H5,H6                         & E                            & -                         \\
                                 &  Hiding Information & Sneak Into Basket$^\dagger$                                    & SN-SI                 & \begin{tabular}[c]{@{}l@{}}\cite{brignull2015dark,gray2018dark,luguri2021shining,Mathur2019,Gray2024,Lacey2023,Saville2024,Yada2022DarkPI}\end{tabular}                            & H1,H2,H5                         & E                            & I/T                       \\
                                 &                                            & \begin{tabular}[c]{@{}l@{}}Drip Pricing, Hidden Costs, Or\\ Partitioned Pricing$^\dagger$\end{tabular}   & SN-DP                 & \begin{tabular}[c]{@{}l@{}}\cite{brignull2015dark,chen2023unveiling,gray2018dark,luguri2021shining,Mathur2019,Gray2024}\\ \cite{Lacey2023,Saville2024,Stavrakakis2021}\end{tabular}                        & H2,H5,H6                         & E                            & I/T                       \\
                                 &                                            & Reference Pricing$^\dagger$                                    & SN-RP                 & \cite{CMA,OECD,Gray2024}                                                                                        & H2,H5,H6                         & E                            & I/T                       \\
                                 &                                            &  Hidden Legalese Stipulations & SN-HL                 & \cite{bosch2016tales,Kennedy2024}                                                                                            & H3,H6                            & ALL                          & T                         \\
                                 & (De)Contextualizing Cues                   & Conflicting Information                              & SN-CI                 & \cite{EDPB,gray2020kind,Gray2024}                                                                                        & H4,H6                            & S                            & T                         \\
                                 &                                            & Information Without Context                          & SN-IW                 & \cite{EDPB,Gray2024}                                                                                            & H4                               & S                            & -                         \\
          &  Fake information*   &  Fake activity*                & SN-FA                 &  \cite{Stavrakakis2021}                                                                       &  H1,H5,H6 &  E    &  T \\ \hline
Interface Interference           & \begin{tabular}[c]{@{}l@{}}Manipulating Visual \\ Choice Architecture\end{tabular}    & False Hierarchy$^\dagger$                                      & II-FH                 & \cite{chen2023unveiling,gray2018dark,luguri2021shining,Gray2024,Lu2024}                                                                                & H2                               & ALL                          & I                         \\
                                 &                                            & Visual Prominence                                    & II-VP                 & \cite{EDPB,Gray2024,Lacey2023}                                                                                         & H1,H2                            & ALL                          & T                         \\
                                 &                                            & Bundling                                             & II-BU                 & \cite{CMA,Gray2024}                                                                                           & H1                               & E                            & I/T                       \\
                                 &                                            & Pressured Selling$^\dagger$                                    & II-PS                 & \cite{FTC,Mathur2019,Gray2024}                                                                                        & H2,H4.H6                         & E                            & I/T                       \\
                                 &                                            & persuasive language*$^\dagger$                                  & II-PL                 & \cite{Mildner2023}                                                                                              & H1,H2,H4                         &  E/S  & -                         \\
                                 & Hard To Close*                            & Small or Moving Close Button*                         & II-SO                 & \cite{FTC}                                                                                               & H4                               & ALL                          & I                         \\
                                 & Bad Defaults/Pre-Selection$^\dagger$                 & -                                                    & II-BD                 & \begin{tabular}[c]{@{}l@{}}\cite{bosch2016tales,chen2023unveiling,gray2018dark,luguri2021shining,Gray2024,Kennedy2024,Lu2024}\end{tabular}                               & H2,H3                            & S                            & I                         \\
                                 & \begin{tabular}[c]{@{}l@{}}Emotional Or Sensory\\ Manipulation\end{tabular}          & Cuteness$^\dagger$                                             & II-CU                 & \cite{Lacey2023,Mathur2019,Gray2024,Stockman2024}                                                                                     &  H1,H2,H5 &  E/ES & -                         \\
                                 &                                            & Positive Or Negative Framing                         & II-PO                 & \cite{chen2023unveiling,Mathur2019,Gray2024}                                                                                       & H2,H3,H4                         & ALL                          & -                         \\
                                 &                                            & Fear Of Missing Out*$^\dagger$                                  & II-FO                 & \cite{Mildner2023}                                                                                               & H4                               & E/S                          & -                         \\
                                 & Trick Questions$^\dagger$                            & -                                                    & II-TQ                 & \begin{tabular}[c]{@{}l@{}}\cite{brignull2015dark,chen2023unveiling,gray2018dark,luguri2021shining,Mathur2019,Gray2024}\\ \cite{Lacey2023,Saville2024,Yada2022DarkPI}\end{tabular}                        & H2,H4,H6                         & ALL                          & I/T                       \\
                                 & Choice Overload                            & -                                                    & II-CO                 & \cite{CMA,EDPB,Gray2024}                                                                                        & H4                               & S                            & I                         \\
                                 &  Hidden Information$^\dagger$ &  -                            & II-HI                 & \cite{gray2018dark,luguri2021shining,Lu2024}                                                                                        & H3,H5,H6                         & ALL                          & I                         \\
                                 & Language Inaccessibility                   & Wrong Language$^\dagger$                                       & II-WL                 & \cite{EDPB,Gray2024}                                                                                            & H3,H4                            & ALL                          & -                         \\
                                 &                                            & Complex Language                                     & II-CL                 & \cite{CMA,Gray2024}                                                                                            & H4                               & ALL                          & -                         \\
                                 & Feedforward Ambiguity                      & -                                                    & II-FA                 & \cite{EDPB,gray2020kind,Gray2024}                                                                                         & H4                               & ALL                          & T                         \\
                                 & Plain Evil*                                 &                                                      & II-PE                 & \cite{Mildner2023}                                                                                              & H1,H4                            & S                            & -                         \\
                                 & Engaging Strategy*                          & Addictive Design*                                     & II-AD                 & \cite{Mildner2023}                                                                                               & H1,H4                            & S                            & -                         \\
                                 &                                            & Infnite Scrolling*                                    & II-IS                 & \cite{Mildner2023}                                                                                               & H1,H4                            & S                            & -                         \\
                                 &                                            & Pull To Refresh*                                      & II-PT                 & \cite{Mildner2023}                                                                                               & H1,H4                            & S                            & -                         \\
                                 &                                            & Reduced Friction*                                     & II-RF                 & \cite{Mildner2023}                                                                                               & H4                               & S                            & -                         \\
                                 &                                            & Customisation*                                        & II-CS                 & \cite{Mildner2023}                                                                                               & H4                               & S                            & -                         \\ \hline
Forced Action                    & Forced Continuity$^\dagger$                          & -                                                    & FA-FC                 & \begin{tabular}[c]{@{}l@{}}\cite{brignull2015dark,chen2023unveiling,gray2018dark,luguri2021shining,Mathur2019,Gray2024}\\ \cite{Lacey2023,Saville2024,Stavrakakis2021}\end{tabular}                        & H1,H2,H5                         & E                            & I/T                       \\
                                 & Forced Registration                        &                                                      & FA-FR                 & \cite{bosch2016tales,luguri2021shining,Mathur2019,Gray2024,Kennedy2024}                                                                                 & H1,H5                            & E/S                          & I/T                       \\
                                 & \begin{tabular}[c]{@{}l@{}}Forced Communication Or\\ Disclosure\end{tabular}         & Privacy Zuckering$^\dagger$                                    & FA-PZ                 & \begin{tabular}[c]{@{}l@{}}\cite{brignull2015dark,chen2023unveiling,gray2018dark,luguri2021shining,Gray2024,Lacey2023}\\ \cite{Kennedy2024,Saville2024,Stavrakakis2021}\end{tabular}                         & H1,H3                            & E/S                          & I                         \\
                                 &                                            & Friend Spam$^\dagger$                                          & FA-FS                 & \cite{brignull2015dark,luguri2021shining,Gray2024,Saville2024,Stavrakakis2021}                                                                                 & H1,H3                            & E/S                          & -                         \\
                                 &                                            & Address Book Leeching$^\dagger$                                & FA-AB                 & \cite{bosch2016tales,luguri2021shining,Gray2024,Kennedy2024}                                                                                    & H1,H3                            & E/S                          & -                         \\
                                 &                                            & Social Pyramid$^\dagger$                                       & FA-SP                 & \cite{chen2023unveiling,gray2018dark,luguri2021shining,Gray2024,Lacey2023}                                                                                 & H1,H3                            & ALL                          & I                         \\
                                 &                                            & Granting and Interaction*                             & FA-GA                 & \cite{Mildner2023}                                                                                               & H1                               & S                            & -                         \\
                                 & Gamification                               & Pay-To-Play$^\dagger$                                          & FA-PT                 & \cite{FTC,gray2018dark,luguri2021shining,Gray2024}                                                                                    & H1,H2                            & ES                           & I                         \\
                                 &                                            & Grinding                                             & FA-GR                 & \cite{FTC,gray2018dark,luguri2021shining,zagal2013dark,Gray2024,Lacey2023}                                                                            & H1,H4                            & ES                           & -                         \\
                                 &                                            & Playing By Appointment*                              & FA-PB                 & \cite{zagal2013dark}                                                                                               & H1                               & ES                           & -                         \\
                                 & Forced Advertisement*                      & Countdown On Ads*                                    & FA-CO                 & \cite{chen2023unveiling}                                                                                               & H4                               & ALL                          & I                         \\
                                 &                                            & \begin{tabular}[c]{@{}l@{}}Watch Ads To Unlock Features\\ Or Get Rewards*\end{tabular}         & FA-WA                 & \cite{chen2023unveiling}                                                                                              & H4                               & ALL                          & I                         \\
                                 &                                            & Pay To Avoid*                                        & FA-PA                 & \cite{chen2023unveiling}                                                                                               & H1,H4                            & ES                           & I/T                       \\
                                 & Automatic Execution*                        & Automating The User Away*                             & FA-AT                 & \cite{gray2020kind}                                                                                               & H1,H4                            & ALL                          & -                         \\
                                 &                                            & Automatic Accept Third Party Term                    & FA-AA                 & \cite{Mildner2023}                                                                                              & H1                               & S                            & -                         \\
                                 & Attention Capture                          & Auto-Play                                            & FA-AP                 & \cite{FTC,Gray2024,Lu2024}                                                                                        & H1,H5                            & ALL                          & I                         \\ \hline
Social Engineering               & Scarcity And Popularity Claims             & High Demand$^\dagger$                                          & SE-HD                 & \cite{luguri2021shining,Mathur2019,Gray2024,Lu2024,Stavrakakis2021}                                                                                  & H2,H6                            & E                            & I/T                       \\
                                 & Social Proof                               & Low Stock$^\dagger$                                            & SE-LS                 & \cite{luguri2021shining,Mathur2019,Gray2024,Stavrakakis2021}                                                                                     & H2,H6                            & E                            & I/T                       \\
                                 &                                            & Endorsement And Testimonials                         & SE-ET                 & \cite{luguri2021shining,Mathur2019,Gray2024}                                                                                        & H2,H6                            & E                            & I/T                       \\
                                 &                                            & Parasocial Pressure$^\dagger$                                  & SE-PP                 & \cite{FTC,luguri2021shining,Mathur2019,Gray2024}                                                                                    & H2,H6                            & E                            & -                         \\
                                 & Urgency                                    & Activity Messages$^\dagger$                                    & SE-AM                 & \cite{luguri2021shining,Mathur2019,Gray2024}                                                                                       & H2,H6                            & E                            & I/T                       \\
                                 &                                            & Countdown Timer                                      & SE-CT                 & \cite{luguri2021shining,Mathur2019,Gray2024}                                                                                        & H2,H6                            & E                            & I/T                       \\
                                 &                                            & Limited Time Message$^\dagger$                                 & SE-LT                 & \cite{luguri2021shining,Mathur2019,Gray2024,Lu2024}                                                                                     & H2,H6                            & E                            & I/T                       \\
                                 & Shaming Personalization                    & Confirmshaming$^\dagger$                                       & SE-CO                 & \cite{brignull2015dark,luguri2021shining,Mathur2019,Gray2024,Saville2024,Stavrakakis2021}                                                                             & H2                               & E                            & I/T                       \\
                                 & Shades Of Grey*                            & Encouraging Anti-Social Behavior*                    & SE-EA                 & \cite{zagal2013dark}                                                                                               & H4                               & ES                           & -                         \\
                                 &                                            & Psychological Tricks*                                & SE-PT                 & \cite{zagal2013dark}                                                                                               & H4,H6                            & ES                           & T                         \\
                                 &                                            & Games For Other Purposes*                            & SE-GF                 & \cite{zagal2013dark}                                                                                               & H4,H6                            & ES                           & -                         \\
                                 & Pre-Delivered Content*$^\dagger$                     & -                                                    & SE-PD                 & \cite{zagal2013dark}                                                                                              & H2                               & ES                           & -                         \\ \hline
\end{tabular}%
}
\begin{minipage}{\textwidth}
\footnotesize
$^\S$: "Common scenarios" refers to the application scenarios in which the dark mode is present. It encompasses four distinct categories: Social Platforms (S), E-commerce (E), Entertainment Software (ES), and General or Universal (All). Dark patterns identified with a '$\dagger$' symbol signify actions or features that result in financial losses and privacy harms for users, highlighting the necessity to prioritize their detection in the realm of dark pattern recognition. Dark patterns labeled with a '*' denote newly introduced dark patterns.

\end{minipage}
\end{table*}
To address Research Question 1 (RQ1), we constructed the most comprehensive and standardized dark pattern taxonomy to date from 113 scientific publications, following the methodology outlined in Section 3.4, as shown in Table\ref{tab:Taxonomy}. This taxonomy covers a total of 68 types of dark patterns (see the third column in Table\ref{tab:Taxonomy}), which fall into 6 main categories (first column) and 32 subcategories (second column). Compared to Gray's taxonomy\cite{gray2023towards}, our taxonomy includes 8 new subcategories and 24 additional types of dark patterns. These newly added patterns are marked with an asterisk in Table\ref{tab:Taxonomy}.
\begin{figure}[h]
\centering
\includegraphics[width=0.55\linewidth]{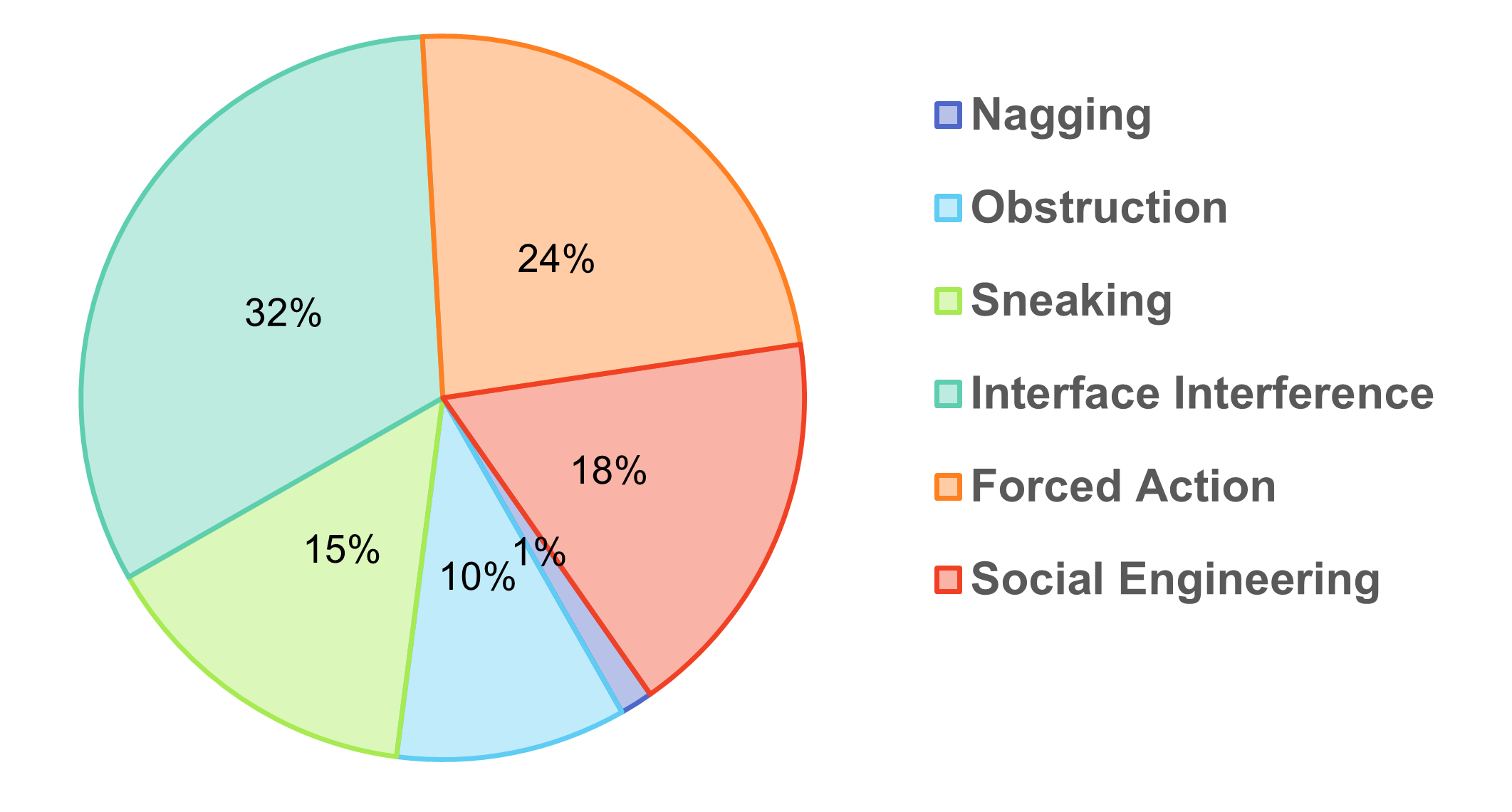}
\vspace{-10pt}
\caption{The proportion of dark pattern types included in different categories}
\vspace{-15pt}
\label{fig:types distribution}
\end{figure}

In Figure \ref{fig:types distribution} , we show the distribution of dark pattern types across each category. It can be seen that among the six categories, “Interface Interference” contains the largest number of dark pattern types, while “Nagging” and “Obstruction” contain relatively fewer. This indicates that existing research has provided more detailed classifications for “Interface Interference,” whereas further exploration is needed to better define subcategories and types within “Nagging” and “Obstruction.”Furthermore, the granularity of the taxonomy can impact the effectiveness of dark pattern detection, highlighting the importance of establishing a comprehensive and well-structured taxonomy for dark patterns. 

The "Severity" column indicates the impact type each dark pattern has on users. Among the identified dark patterns, 49 out of 68 types (72\%) have multi-faceted impacts on users, as evidenced by their association with multiple tags. Among all H1 - H6 impact tags, H2 and H3 emerged as the most severe, as they pertain to financial loss and privacy breaches, respectively. Thirty-seven dark patterns are tagged with these two impact types. If prioritizing dark pattern detection is necessary, patterns tagged with H2 and H3 should be identified first.

The "Common Scenarios" column specifies the potential contexts where each dark pattern might appear. Nineteen dark patterns are commonly found across social platforms (S), e-commerce (E), and entertainment software (ES), suggesting their widespread nature. Additionally, 15 types are unique to social platforms, 17 are specific to e-commerce, and 8 are exclusive to entertainment software, with 7 types appearing in both E and S contexts. Analyzing the potential scenarios for each dark pattern type provides valuable insights for stakeholders, including researchers, designers, and policymakers.

The “Instances” column reveals that 33 types have corresponding image instances, 22 types have corresponding text instances, 17 types have both image and text instances, and 35 types lack associated instances. This observation, based on the  "Instances" column, indicates that dark patterns are primarily studied through image-based data while other data forms remain underexplored. Regarding the incomplete coverage of instances, we attribute this to the following reasons:
\textit{Dataset Limitations}: The accessible datasets have inherent limitations, and our review focused on 113 scientific papers, with limited instances included.
\textit{Complex Interaction Patterns}: Certain dark patterns, such as “Privacy Maze”\cite{EDPB} and “Labyrinthine Navigation”\cite{Mildner2023}, manifest across multiple interaction steps, requiring user interaction across several screens rather than a single page, making it difficult to capture specific instances in one image or text.Detailed instance information can be found in our website.

These findings have received broad support from industry experts. Specifically, as shown in Table ~\ref{table:feedbackresult}, analysis of feedback from 173 valid participants indicates that over 87\% rated our taxonomy of dark patterns as reasonable, comprehensive, understandable, and useful, with scores of 4 or 5. There is general consensus that our annotations for severity and common scenarios are well-justified. Additionally, 10 participants suggested including screenshots for each dark pattern type to enhance understanding, which slightly lowered the approval rate in terms of understandability. Given the extensive manual annotation required, we plan to develop an interactive website for more intuitive dark pattern exploration in future research.
Furthermore, our survey revealed that 61\% of participants were aware of the presence of dark patterns in the applications they developed, 23\% were unaware, and 15\% were uncertain. In terms of mitigation efforts, 30\% reported that their organizations had implemented measures to avoid dark patterns, while 23\% had not, and the remaining 46\% were unsure. This indicates a widespread lack of awareness and proactive measures regarding dark patterns within the industry. More information is available on our website.

\begin{table}[h]
\caption{Feedback Results of Survey}
\scalebox{0.72}{
\begin{tabular}{lcccccll}
\hline
{ \textbf{\# Feedback}} &
  { \textbf{\# Valid feedback}} &
  { \textbf{Rationality}} &
  { \textbf{Completeness}} &
  { \textbf{Understandability}} &
  { \textbf{Usefulness}} &
  \textbf{Severity Annotation} &
  \textbf{Scenarios Annotation} \\ \hline
\multicolumn{1}{c}{{ 176}} &
  { 173} &
  { 100\%} &
  { 92\%} &
  { 87\%} &
  { 92\%} &
  \multicolumn{1}{c}{100\%} &
  \multicolumn{1}{c}{100\%} \\ \hline
\end{tabular}
}
\vspace{45pt}
\label{table:feedbackresult}
\end{table}

\begin{tcolorbox}[]
\vspace{-3pt}
\textbf{Answer to RQ1}:Through the review of 113 scientific papers, we identified and categorized 68 unique types of dark patterns. We further annotated these patterns by category, description, impact on users, and common scenarios. This taxonomy has been validated by industry participants, with over 87\% of feedback indicating that it is reasonable, comprehensive, understandable, and useful. Additionally, we provided instance illustrations for 57\% of the dark pattern types within the taxonomy, enhancing the comprehensibility of different dark pattern types.
\vspace{-3pt}
\end{tcolorbox}
~
\begin{table*}[h]
\centering
\caption{Tool List for Dark Patterns Detection}
\label{tab:ToolsList}
\resizebox{\textwidth}{!}{%
\begin{tabular}{lccllcccc}
\hline
\textbf{Tools} & \textbf{Source}     & \textbf{Year} & \textbf{Technology}                       & \multicolumn{1}{c}{\textbf{Input}} & \textbf{Platform} & \textbf{Auto or Not} & \textbf{Availability} & \textbf{Detected patterns} \\ \hline
AidUI          & ICSE\cite{mansur2023aidui}        & 2023          & Faster-RCNN;heuristic pattern matching    & Image                              & Web \& Mobile     & Auto                 & Yes\cite{toolAidui}            & 11                         \\
UIguard        & UIST\cite{chen2023unveiling}       & 2023          & Faster-RCNN; Knowledge-Driven checker     & Image                              & Mobile            & Auto                 & No                    & 11                         \\
AGM            & PACMHCI\cite{Mathur2019}     & 2019          & Hierarchical Clustering                   & Text                               & Web               & Semi-Auto            & Yes\cite{toolAGM}            & 15                         \\
ADD            & ICDS\cite{curley2021design}        & 2021          & NLP                                       & Text                               & Web               & Auto                 & No                    & 5                          \\
DY             & Mathematics\cite{nazarov2022clustering} & 2022          & Hierarchical and k-means cluster analysis & Text                               & Web               & Auto                 & No                    & 12                         \\
YJT            & BigData\cite{Yada2022DarkPI}     & 2022          & Transformer; BERT; RoBERTa                & Text                               & Web               & Auto                 & Yes\cite{toolYJT}                   & 15                         \\
SMS            & Imaging\cite{Kodandaram2023}    & 2023          & BERT;Transformer;Multi-modal classifier   & Text                               & Web               & Auto                 & No                    & 1                          \\
YMK            & APSEC\cite{Sazid2023}      & 2023          & LLM; Prompt Engineering                   & Text                               & Web               & Auto                 & No                    & 13                         \\ \hline
\end{tabular}%
}
\end{table*}

\subsection{Capability of Detection Tools}

Our goal for RQ2 is to evaluate the capabilities of various dark pattern detection tools based on our comprehensive taxonomy. Following the steps outlined in Section 3.5, we selected eight different dark pattern detection tools: AidUI\cite{mansur2023aidui}, UIGuard\cite{chen2023unveiling}, DY\cite{nazarov2022clustering}, AGM\cite{Mathur2019}, ADD\cite{curley2021design}, SMS\cite{Kodandaram2023}, YMK\cite{Sazid2023}, and YJT\cite{Yada2022DarkPI}. Table\ref{tab:ToolsList} provides an overview of these tools. The first column lists the tools’ names. For unnamed tools, we adopted an abbreviation based on the first letters of the first three authors. Columns 2–3 specify the related conference or journal and publication year. The next column gives an overview of each tool’s main techniques. In the fifth column, we categorize each tool’s input type, followed by a column that identifies the target platform—web or mobile. Columns 7–8 indicate each tool’s automation level and availability, and the last column tallies the number of dark pattern types each tool can detect. Among these tools, six are designed for detecting dark patterns on web pages, one for mobile applications, and another for both platforms. These tools use advanced techniques, including machine learning, natural language processing, and large language models. Except for AGM\cite{Mathur2019}, which requires some human intervention, all the tools claim to be fully automated. Additionally,the three tools—AidUI\cite{mansur2023aidui}, AGM\cite{Mathur2019}, and YJT \cite{Yada2022DarkPI} are open source, while the others are currently unavailable.

We conducted a detailed analysis of these tools based on three aspects: coverage of dark pattern types, performance metrics, and differences in input formats.
~
\begin{table*}[]
\centering
\caption{Analysis of the Capabilities and Limitations of Dark Pattern Detection Tools}
\label{tab:tools}
\resizebox{0.95\linewidth}{!}{%
\begin{tabular}{ccccccccc}
\hline
                                        & \multicolumn{8}{c}{\textbf{Tools(Accuracy,Precison,Recall,F1-score)}}                                                                                                                                                                                                                                                                                                                                                                \\  \cline{2-9} 
                                        & \textbf{AidUI}\cite{mansur2023aidui}                               & \textbf{Uiguard}\cite{chen2023unveiling}                             & \begin{tabular}[c]{@{}c@{}}  \textbf{DY} \cite{nazarov2022clustering}  \end{tabular} & \begin{tabular}[c]{@{}c@{}}\textbf{AGM} \cite{Mathur2019}\end{tabular} & \begin{tabular}[c]{@{}c@{}}\textbf{ADD} \cite{curley2021design}\end{tabular} & \textbf{SMS}\cite{Kodandaram2023}                                    & \textbf{YMK}\cite{Sazid2023}        & \textbf{YJT}\cite{Yada2022DarkPI}     \\  \cline{2-9} 
\multirow{-3}{*}{\textbf{Abbreviation}} & (-,0.66,0.67,0.65)                                   & (-,0.83,0.82,0.82)                                   & -                                                              & -                                                               & -                                                               & (0.86,0.86,0.86,0.86)                                   & -                            & (0.97,0.98,0.96,0.97)    \\ \hline
NG                                      & \checkmark(-,0.52,0.77,0.62)                                  & \checkmark(-,0.67,0.73,0.70)                                  &                                                                &                                                                 &                                                                 &                                                         &                              &                          \\ \hline
OB-IA                                   &                                                      &                                                      & \checkmark                                                              & \checkmark                                                               & \checkmark                                                               &                                                         & \checkmark(-,0.82,0.92,0.86)          & \checkmark                        \\
OB-DE                                   &                                                      &                                                      & \checkmark                                                              & \checkmark                                                               & \checkmark                                                               &                                                         & \checkmark(-,0.82,0.92,0.86)          & \checkmark                        \\
OB-FG                                   &                                                      &                                                      &                                                                &                                                                 &                                                                 &                                                         &                              &                          \\
OB-PC                                   &                                                      &                                                      & \checkmark                                                              &                                                                 &                                                                 &                                                         &                              &                          \\
OB-IC                                   &                                                      &                                                      &                                                                &                                                                 &                                                                 &                                                         &                              &                          \\
OB-PM                                   &                                                      &                                                      &                                                                &                                                                 &                                                                 &                                                         &                              &                          \\
OB-LN                                   &                                                      &                                                      &                                                                &                                                                 &                                                                 &                                                         &                              &                          \\ \hline
SN-DA                                   &   \textbf{\checkmark(-,0.32,0.57,0.41)} &   \textbf{\checkmark(-,0.24,0.50,0.33)} &   \checkmark                                      &                                           &   \checkmark                                       &   \textbf{\checkmark(0.86,0.86,0.86,0.86)} &        &    \\
SN-OI                                   &                                                      &                                                      &                                                                &                                                                 &                                                                 &                                                         &                              &                          \\
SN-CR                                   &                                                      &                                                      &                                                                &                                                                 &                                                                 &                                                         &                              &                          \\
SN-SI                                   &                                                      &                                                      & \checkmark                                                              & \checkmark                                                               &                                                                 &                                                         &                              & \checkmark                        \\
SN-DP                                   &                                                      &                                                      & \checkmark                                                              & \checkmark                                                               &                                                                 &                                                         &                              & \checkmark                        \\
SN-RP                                   &                                                      &                                                      &                                                                &                                                                 &                                                                 &                                                         &                              &                          \\
SN-HL                                   &                                                      &                                                      &                                                                &                                                                 &                                                                 &                                                         &                              &                          \\
SN-CI                                   &                                                      &                                                      &                                                                &                                                                 &                                                                 &                                                         &                              &                          \\
SN-IW                                   &                                                      &                                                      &                                                                &                                                                 &                                                                 &                                                         &                              &                          \\
SN-FA                                   &                                &                                &                                          &                                           &                                           &                                   &        &    \\ \hline
II-FH                                   &                                                      & \checkmark(-,0.65,0.45,0.53)                                  &                                                                & \checkmark                                                               &                                                                 &                                                         &                              & \checkmark                        \\
II-VP                                   &                                                      &                                                      &                                                                &                                                                 &                                                                 &                                                         & \checkmark(-,0.97,0.65,0.78)          &                          \\
II-BU                                   &                                                      &                                                      &                                                                &                                                                 &                                                                 &                                                         &                              &                          \\
II-PS                                   &                                                      &                                                      &                                                                & \checkmark                                                               &                                                                 &                                                         & \checkmark(-,0.97,0.65,0.78)          & \checkmark                        \\
II-PL                                   &                                                      &                                                      &                                                                &                                                                 &                                                                 &                                                         &                              &                          \\
II-SO                                   &                                                      & \textbf{\checkmark(-,0.99,0.94,0.97)}                         &                                                                &                                                                 &                                                                 &                                                         &                              &                          \\
II-BD                                   & \checkmark(-,0.67,0.59,0.62)                                  & \checkmark(-,0.86,0.75,0.80)                                  &                                                                &                                                                 &                                                                 &                                                         &                              &                          \\
II-CU                                   &                                                      &                                                      &                                                                &                                                                 &                                                                 &                                                         &                              &                          \\
II-PO                                   &                                                      &                                                      &                                                                &                                                                 &                                                                 &                                                         &                              &                          \\
II-FO                                   &                                                      &                                                      &                                                                &                                                                 &                                                                 &                                                         &                              &                          \\
II-TQ                                   &                                                      &                                                      & \checkmark                                                              & \checkmark                                                               & \checkmark                                                               &                                                         & \checkmark(-,0.97,0.65,0.78)          & \textbf{\checkmark}               \\
II-CO                                   &                                                      &                                                      &                                                                &                                                                 &                                                                 &                                                         &                              &                          \\
II-HI                                   &                                                      &                                                      &                                                                &                                                                 &                                                                 &                                                         &                              &                          \\
II-WL                                   &                                                      &                                                      &                                                                &                                                                 &                                                                 &                                                         &                              &                          \\
II-CL                                   &                                                      &                                                      &                                                                &                                                                 &                                                                 &                                                         &                              &                          \\
II-FA                                   &                                                      &                                                      &                                                                &                                                                 &                                                                 &                                                         &                              &                          \\
II-PE                                   &                                                      &                                                      &                                                                &                                                                 &                                                                 &                                                         &                              &                          \\
II-AD                                   &                                                      &                                                      &                                                                &                                                                 &                                                                 &                                                         &                              &                          \\
II-IS                                   &                                                      &                                                      &                                                                &                                                                 &                                                                 &                                                         &                              &                          \\
II-PT                                   &                                                      &                                                      &                                                                &                                                                 &                                                                 &                                                         &                              &                          \\
II-RF                                   &                                                      &                                                      &                                                                &                                                                 &                                                                 &                                                         &                              &                          \\
II-CS                                   &                                                      &                                                      &                                                                &                                                                 &                                                                 &                                                         &                              &                          \\ \hline
FA-FC                                   &                                                      & \checkmark(-,1.0,1.0,1.0)                                     & \checkmark                                                              &                                                                 &                                                                 &                                                         &                              &                          \\
FA-FR                                   &                                                      &                                                      &                                                                & \checkmark                                                               &                                                                 &                                                         & \textbf{\checkmark(-,0.04,1.0,0.08)}  & \checkmark                        \\
FA-PZ                                   &                                                      & \checkmark(-,0.80,0.67,0.73)                                  & \checkmark                                                              &                                                                 &                                                                 &                                                         &                              &                          \\
FA-FS                                   &                                                      &                                                      & \checkmark                                                              &                                                                 & \checkmark                                                               &                                                         &                              &                          \\
FA-AB                                   &                                                      &                                                      &                                                                &                                                                 &                                                                 &                                                         &                              &                          \\
FA-SP                                   &                                                      & \checkmark(-,0.50,0.88,0.64)                                  &                                                                &                                                                 &                                                                 &                                                         &                              &                          \\
FA-GA                                   &                                                      &                                                      &                                                                &                                                                 &                                                                 &                                                         &                              &                          \\
FA-PT                                   & \checkmark(-,0.80,0.36,0.50)                                  &                                                      &                                                                &                                                                 &                                                                 &                                                         &                              &                          \\
FA-GR                                   & \checkmark(-,0.80,0.36,0.50)                                  &                                                      &                                                                &                                                                 &                                                                 &                                                         &                              &                          \\
FA-PB                                   &                                                      &                                                      &                                                                &                                                                 &                                                                 &                                                         &                              &                          \\
FA-CO                                   &                                                      & \checkmark(-,0.75,0.75,0.75)                                  &                                                                &                                                                 &                                                                 &                                                         &                              &                          \\
FA-WA                                   &                                                      & \checkmark(-,0.67,1.0,0.80)                                   &                                                                &                                                                 &                                                                 &                                                         &                              &                          \\
FA-PA                                   &                                                      & \checkmark(-,0.87,0.90,0.88)                                  &                                                                &                                                                 &                                                                 &                                                         &                              &                          \\
FA-AT                                   &                                                      &                                                      &                                                                &                                                                 &                                                                 &                                                         &                              &                          \\
FA-AA                                   &                                                      &                                                      &                                                                &                                                                 &                                                                 &                                                         &                              &                          \\
FA-AP                                   & \checkmark(-,0.55,0.46,0.50)                                  &                                                      & \checkmark                                                              &                                                                 &                                                                 &                                                         &                              &                          \\ \hline
SE-HD                                   & \textbf{\checkmark(-,1.0,0.80,0.89)}                          &                                                      &                                                                & \checkmark                                                               &                                                                 &                                                         & \checkmark(-,0.92,0.98,0.95)          & \checkmark                        \\
SE-LS                                   & \checkmark-,0.75,0.79,0.77)                                   &                                                      &                                                                & \checkmark                                                               &                                                                 &                                                         & \textbf{\checkmark(-,0.99,0.94,0.97)} & \checkmark                        \\
SE-ET                                   &                                                      &                                                      &                                                                & \checkmark                                                               &                                                                 &                                                         & \textbf{\checkmark(-,0.99,0.94,0.97)} & \checkmark                        \\
SE-PP                                   &                                                      &                                                      &                                                                &                                                                 &                                                                 &                                                         &                              &                          \\
SE-AM                                   & \textbf{\checkmark(1.0,0.80,0.89)}                            &                                                      &                                                                & \checkmark                                                               &                                                                 &                                                         & \checkmark(-,0.90,0.93,0.91)          & \checkmark                        \\
SE-CT                                   & \checkmark(0.63,0.86,0.73)                                    &                                                      &                                                                & \checkmark                                                               &                                                                 &                                                         & \checkmark(-,0.90,0.93,0.91)          & \checkmark                        \\
SE-LT                                   & \checkmark(0.95,0.73,0.83)                                    &                                                      &                                                                & \checkmark                                                               &                                                                 &                                                         & \checkmark(-,0.90,0.93,0.91)          & \checkmark                        \\
SE-CO                                   &                                                      &                                                      & \checkmark                                                              & \checkmark                                                               &                                                                 &                                                         & \checkmark(-,0.97,0.65,0.78)          & \checkmark                        \\
SE-EA                                   &                                                      &                                                      &                                                                &                                                                 &                                                                 &                                                         &                              &                          \\
SE-PT                                   &                                                      &                                                      &                                                                &                                                                 &                                                                 &                                                         &                              &                          \\
SE-GF                                   &                                                      &                                                      &                                                                &                                                                 &                                                                 &                                                         &                              &                          \\
SE-PD                                   & \multicolumn{1}{l}{}                                 & \multicolumn{1}{l}{}                                 & \multicolumn{1}{l}{}                                           & \multicolumn{1}{l}{}                                            & \multicolumn{1}{l}{}                                            & \multicolumn{1}{l}{}                                    & \multicolumn{1}{l}{}         & \multicolumn{1}{l}{}     \\ \hline
\end{tabular}%
}
\end{table*}

\textit{Analysis of Type Coverage}: We first analyzed the coverage of dark pattern types by each tool. Table\ref{tab:tools} explores these tools' detection capabilities against our taxonomy. Columns 2–9 represent each tool, with a checkmark indicating if a tool can detect a particular dark pattern. According to Table \ref{tab:tools}, these eight tools collectively detected 31 different dark patterns, achieving a coverage rate of 45.5\% (31 out of 68). Most detectable dark patterns are recognized by two or more tools. Specifically, nine patterns are detected by only one tool, while 22 are detected by two or more tools. Looking at each tool individually, the number of detectable dark patterns varies from one to 15. AGM\cite{Mathur2019} and YJT\cite{Yada2022DarkPI} detect the most patterns at 15 each, while SMS\cite{Kodandaram2023} detects only one pattern. Notably, the tool with the highest type coverage still detects only 22\% (15 out of 68), indicating that the current tools’ coverage of dark patterns remains relatively limited. Importantly, 37 dark patterns (54.5\% of the total) are undetected by any tool. Several factors may explain why these patterns are undetected: some patterns are inherently complex, making detection challenging; others manifest in multiple ways, complicating clear detection rules; and some appear infrequently, leading to a lack of representative samples for comprehensive understanding and detection. These findings highlight significant limitations in existing dark pattern detection tools, underscoring a gap between the range of dark patterns identified in our taxonomy and those detectable by current tools. This emphasizes the need for more advanced and comprehensive tools to protect users from a broader range of dark patterns.

\textit{Performance Metrics Analysis}: In Table \ref{tab:tools}, we provide detailed overall performance metrics and metrics for detecting specific types across tools. We conducted both horizontal and vertical comparisons.

Horizontal Comparison: When examining overall performance metrics, we find that YJT\cite{Yada2022DarkPI} performs best, with metrics (0.97, 0.98, 0.96, 0.97), while AidUI\cite{mansur2023aidui} has comparatively lower metrics (-, 0.66, 0.67, 0.65).

Vertical Comparison: This approach mainly assesses the variation in detection effectiveness across types within each tool. We set 0.5 as the threshold for judging a tool's effectiveness in detecting a particular type. Metrics significantly below 0.5 indicate poor detection, while those significantly above 0.5 indicate good detection. Based on Table\ref{tab:tools}, AidUI\cite{mansur2023aidui} performs best in detecting “High Demand” and “Activity Message” types. UIGuard\cite{chen2023unveiling} is most effective in detecting “Small or Moving Close Button” (although some metrics for “Forced Continuity” are 1, we did not consider it the best type for this tool due to limited instances). SMS\cite{Kodandaram2023} can detect only one dark pattern type, “Disguised Ad,” which has the poorest detection metrics in both AidUI\cite{mansur2023aidui} and UIGuard\cite{chen2023unveiling}. YMK\cite{Sazid2023} shows the highest metrics for “Low Stock” and “Endorsement And Testimonials,” with the lowest metrics for “Forced Registration.” 

This metric comparison reveals that while current tools can detect a limited number of dark pattern types, their performance varies significantly across types, with some types performing poorly. This suggests that further optimization and adjustment of the models used by current tools is needed to enhance detection accuracy progressively.

\textit{Input Type Analysis}: The eighth column of Table \ref{tab:ToolsList} records each tool's input type. Tools with image input types include AidUI\cite{mansur2023aidui} and UIGuard\cite{chen2023unveiling}, while those with text input types include YMK\cite{Sazid2023}, YJT\cite{Yada2022DarkPI}, ADD\cite{curley2021design}, AGM\cite{Mathur2019}, SMS\cite{Kodandaram2023}, and DY\cite{nazarov2022clustering}. Based on Table\ref{tab:tools} annotations, we can see that image-input tools can detect a total of 19 types, while text-input tools detect a total of 22 types, with 10 types detectable by both input methods, nine types detectable only by image-input tools, and 12 detectable only by text-input tools. Although text-input tools detect a broader range of types, there are still nine types that text-input tools cannot detect. The detection differences between input formats highlight the diversity of dark pattern expressions and suggest that the handling of different data forms in dark pattern detection tools requires further research to expand detection coverage.

Through this three-pronged analysis, we observe that these tools employ various technical approaches and focus on different aspects of research, with significant variation in the dark pattern types they can recognize. Analysis of detection coverage, type performance metrics, and input format differences highlights both the capabilities and limitations of existing detection tools.

\vspace{-5pt}
\begin{tcolorbox}[]
\vspace{-3pt}
\textbf{Answer to RQ2}:  
Our assessment of detection tools reveals that among the eight tools, only three are open-source, and their combined detection coverage is limited, identifying only 31 out of the 68 types in our taxonomy, resulting in a 45.5\% coverage rate. However, 37 dark pattern types (54.5\% of the total) remain undetected by any tool. Additionally, performance metrics for some types fall below 50\%, and detection effectiveness varies across input formats. These limitations emphasize the current detection tools’ restricted coverage and suggest that further advancements are needed to address the diverse range of dark patterns. 
\vspace{-3pt}
\end{tcolorbox}
\vspace{-5pt}

\begin{table*}[h]
\centering
\caption{Dataset List for Dark Patterns}
\label{tab:DatasetsList}
\resizebox{\linewidth}{!}{%
\begin{tabular}{lccccccc}
\hline
\textbf{Dataset} & \textbf{C/J} & \textbf{Year} & \textbf{Type} & \textbf{Platform} & \textbf{\#Instances} & \textbf{\#Covered Patterns} & \textbf{Availability} \\ \hline
ContextDP-D      & ICSE\cite{mansur2023aidui} & 2023          & Image         & Web \& Mobile     & 301                  & 10                          & Yes\cite{urlContextDP}            \\
Rico’-D          & UIST\cite{chen2023unveiling} & 2023          & Image         & Mobile            & 1,660                & 11                          & Yes\cite{urlRico}            \\
AGM-D            & HCI\cite{Mathur2019}  & 2019          & Text          & Web               & 1,818                & 14                          & Yes\cite{urlAGM}            \\
LLE-D            & CHI\cite{di2020ui}  & 2020          & Video         & Mobile            & 1,782                & 18                          & Yes(partial)\cite{urlLLE}   \\ \hline
\end{tabular}%
}
\end{table*}
\begin{table*}[]
\centering
\caption{Analysis of the Coverage of Dark Pattern-Related Datasets}
\label{tab:datasets}
\resizebox{0.75\linewidth}{!}{%
\begin{tabular}{cllcllcllcllcll}
\hline
                                        & \multicolumn{14}{c}{\textbf{Datasets(sample size)}}                                                                                                                         \\ \cline{2-15} 
\multirow{-2}{*}{\textbf{Abbreviation}} &  &  & \textbf{ContextDP-D (301)}\cite{urlContextDP} &  &  & \textbf{Rico’-D (1660)}\cite{urlRico} &  &  & \textbf{AGM-D (1818)}\cite{urlAGM} &  &  & \textbf{LLE-D (1782)}\cite{urlLLE}   &  &  \\ \hline
NG                                      &  &  & \checkmark(57)                             &  &  & \checkmark(188)                          &  &  &                               &  &  & \checkmark(352)                         &  &  \\ \hline
OB-IA                                   &  &  &                                   &  &  &                                 &  &  & \checkmark(31)                         &  &  & \checkmark(110)                         &  &  \\
OB-DE                                   &  &  &                                   &  &  &                                 &  &  &                               &  &  &                                &  &  \\
OB-FG                                   &  &  &                                   &  &  &                                 &  &  &                               &  &  &                                &  &  \\
OB-PC                                   &  &  &                                   &  &  &                                 &  &  &                               &  &  & \checkmark(25)                          &  &  \\
OB-IC                                   &  &  &                                   &  &  &                                 &  &  &                               &  &  & \checkmark(23)                          &  &  \\
OB-PM                                   &  &  &                                   &  &  &                                 &  &  &                               &  &  &                                &  &  \\
OB-LN                                   &  &  &                                   &  &  &                                 &  &  &                               &  &  &                                &  &  \\ \hline
SN-DA                                   &  &  &    \checkmark(21)     &  &  &    \checkmark(46)   &  &  &          &  &  &    \checkmark(184) &  &  \\
SN-OI                                   &  &  &                                   &  &  &                                 &  &  &                               &  &  &                                &  &  \\
SN-CR                                   &  &  &                                   &  &  &                                 &  &  &                               &  &  &                                &  &  \\
SN-SI                                   &  &  &                                   &  &  &                                 &  &  & \checkmark(21)                         &  &  & \checkmark(3)                           &  &  \\
SN-DP                                   &  &  &                                   &  &  &                                 &  &  & \checkmark(5)                          &  &  &                                &  &  \\
SN-RP                                   &  &  &                                   &  &  &                                 &  &  &                               &  &  &                                &  &  \\
SN-HL                                   &  &  &                                   &  &  &                                 &  &  &                               &  &  &                                &  &  \\
SN-CI                                   &  &  &                                   &  &  &                                 &  &  &                               &  &  &                                &  &  \\
SN-IW                                   &  &  &                                   &  &  &                                 &  &  &                               &  &  &                                &  &  \\
SN-FA                                   &  &  &              &  &  &            &  &  &          &  &  &           &  &  \\ \hline
II-FH                                   &  &  &                                   &  &  & \checkmark(273)                          &  &  &                               &  &  & \checkmark(299)                         &  &  \\
II-VP                                   &  &  &                                   &  &  &                                 &  &  & \checkmark(25)                         &  &  &                                &  &  \\
II-BU                                   &  &  &                                   &  &  &                                 &  &  &                               &  &  &                                &  &  \\
II-PS                                   &  &  &                                   &  &  &                                 &  &  & \checkmark(67)                         &  &  &                                &  &  \\
II-PL                                   &  &  &                                   &  &  &                                 &  &  &                               &  &  &                                &  &  \\
II-SO                                   &  &  &                                   &  &  & \checkmark(684)                          &  &  &                               &  &  & \checkmark(114)                         &  &  \\
II-BD                                   &  &  & \checkmark(111)                            &  &  & \checkmark(232)                          &  &  &                               &  &  & \checkmark(210)                         &  &  \\
II-CU                                   &  &  &                                   &  &  &                                 &  &  &                               &  &  &                                &  &  \\
II-PO                                   &  &  &                                   &  &  &                                 &  &  &                               &  &  &                                &  &  \\
II-FO                                   &  &  &                                   &  &  &                                 &  &  &                               &  &  &                                &  &  \\
II-TQ                                   &  &  & \textbf{}                         &  &  & \textbf{}                       &  &  & \checkmark(9)                          &  &  & \checkmark(1)                           &  &  \\
II-CO                                   &  &  &                                   &  &  &                                 &  &  &                               &  &  &                                &  &  \\
II-HI                                   &  &  &                                   &  &  &                                 &  &  &                               &  &  & \checkmark(146)                         &  &  \\
II-WL                                   &  &  &                                   &  &  &                                 &  &  &                               &  &  &                                &  &  \\
II-CL                                   &  &  &                                   &  &  &                                 &  &  &                               &  &  &                                &  &  \\
II-FA                                   &  &  &                                   &  &  &                                 &  &  &                               &  &  &                                &  &  \\
II-PE                                   &  &  &                                   &  &  &                                 &  &  &                               &  &  &                                &  &  \\
II-AD                                   &  &  &                                   &  &  &                                 &  &  &                               &  &  &                                &  &  \\
II-IS                                   &  &  &                                   &  &  &                                 &  &  &                               &  &  &                                &  &  \\
II-PT                                   &  &  &                                   &  &  &                                 &  &  &                               &  &  &                                &  &  \\
II-RF                                   &  &  &                                   &  &  &                                 &  &  &                               &  &  &                                &  &  \\
II-CS                                   &  &  &                                   &  &  &                                 &  &  &                               &  &  &                                &  &  \\ \hline
FA-FC                                   &  &  &                                   &  &  & \checkmark(1)                            &  &  &                               &  &  & \checkmark(48)                          &  &  \\
FA-FR                                   &  &  &                                   &  &  &                                 &  &  & \checkmark(6)                          &  &  &                                &  &  \\
FA-PZ                                   &  &  &                                   &  &  & \checkmark(117)                          &  &  &                               &  &  & \checkmark(87)                          &  &  \\
FA-FS                                   &  &  &                                   &  &  &                                 &  &  &                               &  &  &                                &  &  \\
FA-AB                                   &  &  &                                   &  &  &                                 &  &  &                               &  &  &                                &  &  \\
FA-SP                                   &  &  &                                   &  &  & \checkmark(16)                           &  &  &                               &  &  & \checkmark(15)                          &  &  \\
FA-GA                                   &  &  &                                   &  &  &                                 &  &  &                               &  &  &                                &  &  \\
FA-PT                                   &  &  & \checkmark(11)                             &  &  &                                 &  &  &                               &  &  &                                &  &  \\
FA-GR                                   &  &  &                                   &  &  &                                 &  &  &                               &  &  &                                &  &  \\
FA-PB                                   &  &  &                                   &  &  &                                 &  &  &                               &  &  &                                &  &  \\
FA-CO                                   &  &  &                                   &  &  & \checkmark(4)                            &  &  &                               &  &  & \checkmark(72)                          &  &  \\
FA-WA                                   &  &  &                                   &  &  & \checkmark(2)                            &  &  &                               &  &  & \checkmark(50)                          &  &  \\
FA-PA                                   &  &  &                                   &  &  & \checkmark(97)                           &  &  &                               &  &  &                                &  &  \\
FA-AT                                   &  &  &                                   &  &  &                                 &  &  &                               &  &  &                                &  &  \\
FA-AA                                   &  &  &                                   &  &  &                                 &  &  &                               &  &  &                                &  &  \\
FA-AP                                   &  &  & \checkmark(13)                             &  &  &                                 &  &  &                               &  &  &                                &  &  \\ \hline
SE-HD                                   &  &  & \checkmark(5)                              &  &  &                                 &  &  & \checkmark(47)                         &  &  &                                &  &  \\
SE-LS                                   &  &  & \checkmark(19)                             &  &  &                                 &  &  & \checkmark(632)                        &  &  &                                &  &  \\
SE-ET                                   &  &  &                                   &  &  &                                 &  &  & \checkmark(12)                         &  &  &                                &  &  \\
SE-PP                                   &  &  &                                   &  &  &                                 &  &  &                               &  &  &                                &  &  \\
SE-AM                                   &  &  & \checkmark(10)                             &  &  &                                 &  &  & \checkmark(313)                        &  &  &                                &  &  \\
SE-CT                                   &  &  & \checkmark(28)                             &  &  &                                 &  &  & \checkmark(393)                        &  &  & \checkmark(30)                          &  &  \\
SE-LT                                   &  &  & \checkmark(26)                             &  &  &                                 &  &  & \checkmark(88)                         &  &  &                                &  &  \\
SE-CO                                   &  &  &                                   &  &  &                                 &  &  & \checkmark(169)                        &  &  & \checkmark(13)                          &  &  \\
SE-EA                                   &  &  &                                   &  &  &                                 &  &  &                               &  &  &                                &  &  \\
SE-PT                                   &  &  &                                   &  &  &                                 &  &  &                               &  &  &                                &  &  \\
SE-GF                                   &  &  &                                   &  &  &                                 &  &  &                               &  &  &                                &  &  \\
SE-PD                                   &  &  &                                   &  &  &                                 &  &  &                               &  &  &                                &  &  \\ \hline
\end{tabular}%
}
\end{table*}
\subsection{Availability of Datasets}
\subsubsection{Dataset status}
For RQ3, we analyzed the datasets in terms of instance types, dataset scale, and the coverage of dark pattern types to assess their comprehensiveness and completeness.

As described in Section 3.6, we selected a total of four datasets containing instances of dark patterns: ContextDP-D\cite{mansur2023aidui}, Rico’-D\cite{chen2023unveiling}, AGM-D \cite{Mathur2019}, and LLE-D\cite{di2020ui}. Table\ref{tab:DatasetsList} provides detailed information on these datasets. The first column describes the dataset's name. Notably, if a dataset does not have a designated name provided by the authors, we established a unified naming convention: using the initials of the first three authors followed by a “-D” suffix. Next are the conference or journal of the related paper and the year of publication. The fourth column presents the format of instances included in each dataset, while the fifth column specifies the target platform, whether web or mobile. The sixth and seventh columns describe the dataset’s scale (the number of instances it contains) and the number of dark pattern types it covers. The final column highlights the availability of each dataset, with a direct link included if applicable. As shown in Table\ref{tab:DatasetsList}, all datasets except for LLE-D\cite{di2020ui}, which only partially provides its data, are directly accessible. Based on the information in Table\ref{tab:DatasetsList} and \ref{tab:datasets}, we conducted the following analysis.

Regarding instance types, among the four datasets, two are image-based, while the other two are text-based and video-based, respectively. Overall, the number of currently available datasets is relatively low, and the instance types within these datasets are limited, with image-based types predominating. This lack of multimodal datasets restricts the capability to detect diverse dark patterns, potentially limiting the generalization of tools and reducing detection effectiveness in complex scenarios.

Regarding dataset scale, the four datasets contain a total of 5,561 instances. More specifically, Rico’-D and LLE-D each contain 1,660 and 1,782 dark pattern instances from mobile applications, respectively. AGM-D includes 1,818 instances from webpages, while ContextDP-D contains fewer instances, with 301 from both platforms.

\begin{figure}[h]
\centering
\includegraphics[width=1\textwidth]{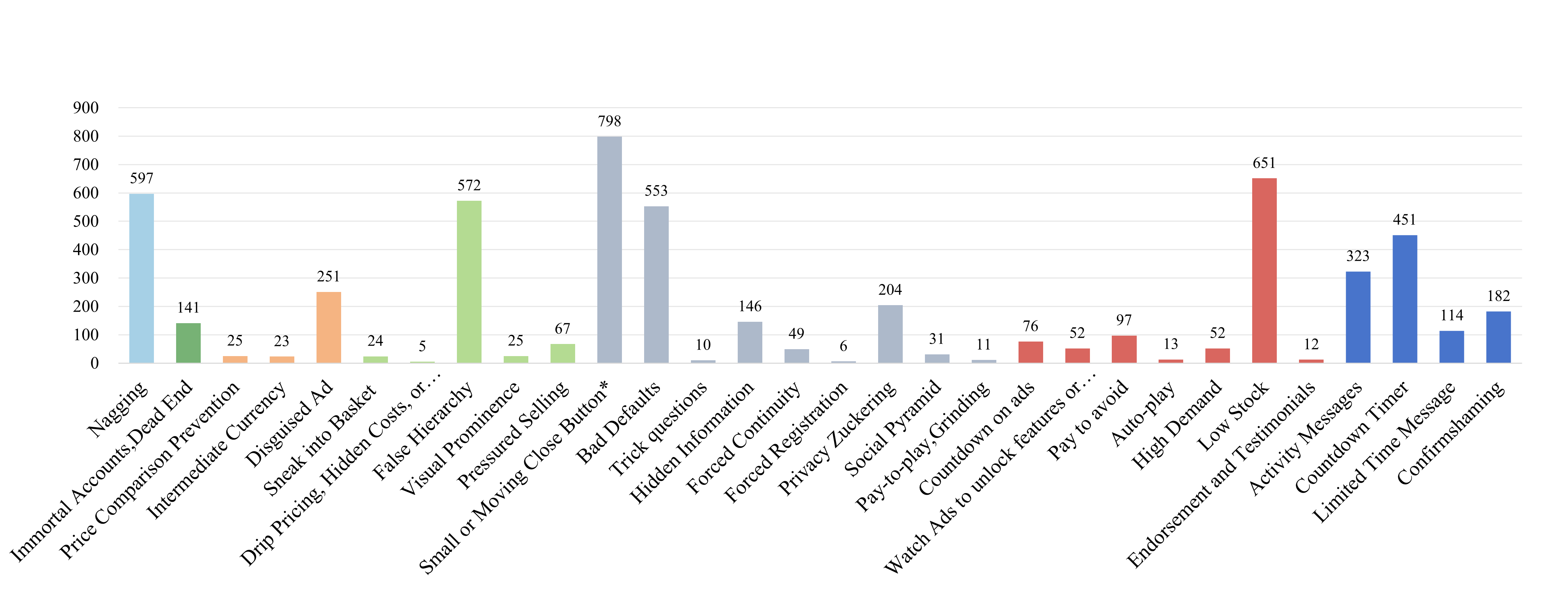}
\vspace{-5pt}
\caption{Comprehensive dark pattern instances count across three datasets}
\vspace{-5pt}
\label{fig:instances}
\end{figure}

Regarding dataset coverage, as shown in Table\ref{tab:datasets}, these four datasets collectively cover only 30 dark patterns, with an overall coverage rate of 44\% (= 30/68). When considering each dataset individually, the range of covered dark patterns further narrows, capturing only between 10 and 18 different types. Additionally, as illustrated in Figure \ref{fig:instances}, the number of instances for each dark pattern within the datasets varies significantly. For instance, “Low Stock” and “Small or Moving Close Button” have notably high instance counts, each exceeding 600. This suggests that these dark patterns may be more common in daily applications, making their instances easier to collect. In contrast, six dark patterns, such as “Drip Pricing, Hidden Costs, Or Partitioned Pricing” and “Forced Registration,” have much lower instance counts, with fewer than 10 instances each. This pronounced data imbalance may jeopardize the performance of machine learning-based dark pattern prediction and classification, potentially introducing model bias that affects accuracy and robustness.

Moreover, the absence of 38 dark patterns (56\% of the total) from the datasets indicates that machine learning models may fail to capture these patterns’ features, resulting in ineffective recognition and classification for such instances. Possible reasons for these missing dark patterns are as follows: (1) Ambiguity and lack of precise standards: Some dark patterns are vaguely defined, which often hinders researchers from clearly identifying them in screenshots of mobile applications or webpages. For example, although “complex language” is described as “decision-related information that may be intentionally or unintentionally made difficult to understand through the use of complex language”\cite{CMA,oppenheimer2006consequences}, there is no concrete benchmark defining what constitutes complex language. This lack of clarity may mislead researchers attempting to identify such patterns. (2) Dependence on context and historical user interaction: Some dark patterns are not limited to a single interface but manifest based on the user’s interaction history or broader context. Thus, they are difficult to detect immediately in isolated interfaces, making data collection challenging. A relevant example is “Address Book Leeching,” where a service conducts a database search when a user imports their contact list. Although these services may store these contact details for undisclosed reasons not initially revealed to the user, malicious intent may only become apparent in subsequent interactions\cite{bosch2016tales}. Current methods of collecting dark pattern instances mainly rely on manually identifying and labeling screenshots or videos of mobile applications or webpages, lacking the capability to recognize such subtle patterns based on a single interface.

Our analysis of these datasets has significant implications for dark pattern detection using machine learning techniques. With a pronounced data imbalance where certain patterns are overrepresented and others are underrepresented or even missing, models trained on these datasets may show higher sensitivity to more common patterns, while their ability to recognize rarer patterns may be diminished. Additionally, the complete absence of 38 dark pattern types highlights a substantial gap in current datasets, limiting models’ ability to detect a broad range of dark patterns. Therefore, to achieve reliable and comprehensive dark pattern detection, there is an urgent need for more balanced and detailed datasets that cover the various dark patterns encountered in real-world scenarios.
\subsubsection{Standard Dataset}
Based on the taxonomy established in this paper, we constructed a standard dataset and aligned the categorization of available datasets to maintain consistency with our taxonomy.

As described in Section 3.7, we processed three datasets (ContextDP-D\cite{mansur2023aidui}, Rico'-D\cite{chen2023unveiling}, and AGM-D\cite{Mathur2019}) accordingly, while LLE-D\cite{di2020ui} was not processed due to incomplete data. Through this method, we obtained a standardized text dataset containing 1,511 dark pattern instances and an image dataset with 1,822 instances, both aligned to our taxonomy. 

The text dataset was created by processing AGM-D\cite{Mathur2019}, where we removed 307 duplicate and blank entries, and updated 15 type labels to match the categories in our taxonomy. For the image dataset, we updated 10 type labels in ContextDP-D\cite{mansur2023aidui} and 11 in Rico'-D\cite{chen2023unveiling} to correspond to our taxonomy. Since the format of instances in the two image datasets was the same, we merged them. As ContextDP-D\cite{mansur2023aidui} images from the Mobile platform overlapped with those in Rico'-D\cite{chen2023unveiling}, we combined ContextDP-D's 162 Web instances with Rico'-D's 1,660 instances, forming an image dataset containing 1,822 instances with unified labels. This merged dataset covers more types, which will be beneficial for future research. The processed data will be made available on our website for reference in subsequent studies.

\vspace{-5pt}
\begin{tcolorbox}[]
\vspace{-3pt}
\textbf{Answer to RQ3}: 
These four datasets contain a total of 5,561 instances and cover only 30 different dark pattern types, yielding a total coverage rate of 44\% (30 out of 68 types). The lack of 38 types (56\% of the total) limits the ability to identify these patterns. Through the construction of this standardized dataset, we obtained a labeled image dataset with 1,822 dark pattern instances covering 18 types, and a text dataset with 1,511 instances covering 15 types. These newly standardized datasets will facilitate future research by providing a unified categorization framework.
\vspace{-3pt}
\end{tcolorbox}
\vspace{-5pt}

\section{Discussion} 

\subsection{Implications}
We have developed a comprehensive taxonomy of dark patterns, which has significant implications for both academia and industry:
\begin{itemize}
    \item \textit{Standardization}: For researchers, developers, and policymakers, this taxonomy establishes a unified nomenclature that reduces ambiguity, facilitates interdisciplinary collaboration, and promotes knowledge sharing and technological advancement across related fields.
    \item \textit{Impact Assessment of Dark Patterns}: This taxonomy aids developers and policymakers by categorizing dark patterns based on both the level of harm to users and their usage contexts. This practical framework informs design decisions and policy interventions, enabling more scientific and effective measures in policy-making.
    \item \textit{Dark Pattern Instance Visualization}:  For researchers, developers, and policymakers, we have collected and visualized instances of various types of dark patterns based on our taxonomy. This visualization not only enhances understanding but also supports effective differentiation between types, advancing in-depth research.
    \item \textit{Evaluation of Detection Tools and Available Datasets}: In this paper, we conducted an in-depth evaluation of existing detection tools and datasets. This assessment highlights the strengths and limitations of current detection tools and discusses the availability of datasets, providing a reference for developing new detection tools and datasets.
    \item \textit{Construction of a Standard Dataset}:Building on existing datasets, we made adjustments and constructed a new standard dataset according to the taxonomy proposed in this paper. This dataset provides a solid foundation for future research, supporting further exploration and innovation by researchers and developers.
\end{itemize}

\subsection{Threats to Validity}
\subsubsection{Internal Threats}

The manual annotation process used to identify user impact dimensions and application scenarios may introduce subjectivity and potential inaccuracies. Although we mitigated this risk through expert consultations, human error cannot be completely eliminated. This study relies on an industry survey to validate the taxonomy, and the quality of this validation largely depends on respondents' understanding and interpretation of the taxonomy, which introduces an additional layer of subjectivity. Furthermore, the methods used to evaluate existing dark pattern detection tools may not cover all possible configurations and nuances, potentially impacting the comprehensiveness of our evaluation.

\subsubsection{External Threats}

First, our validation relies on a survey conducted with industry professionals, but the sample may not fully represent the diverse range of practitioners in this field. Second, the study primarily focuses on industry professionals familiar with Western design paradigms; thus, the taxonomy and its implications may not be universally applicable across different cultural and geographical contexts. Third, we assessed dark pattern detection tools based on their current capabilities. As these tools evolve and improve, the relevance of our evaluation may change over time. Fourth, our evaluation of datasets reflects their current state, yet as datasets become richer and more refined through continued research, our findings may need to be updated accordingly.

\subsection{Challenges}
Research on dark patterns presents inherent challenges due to the complexity of accuracy in algorithms, data completeness, and tool usability. Our study provides a foundational framework to address these challenges while highlighting multiple avenues and challenges for future research.
\begin{itemize}
    \item \textit{Data Quality and Quantity}: The robustness of the classification system depends on data quality and quantity. Although we have strived to ensure data accuracy through expert consultations, the manual labeling process remains vulnerable to errors and subjectivity.
    \item \textit{Development of Open-Source Tools}: Developing open-source tools for detecting dark patterns is essential but poses evaluation challenges. Current methods may struggle to capture the full diversity and subtlety of dark patterns, risking biases that could affect their general applicability.
    \item \textit{Rich and Balanced Datasets}:  There is an urgent need for datasets that are diverse and internally balanced. Existing datasets do not fully meet the needs of in-depth research, so future efforts should employ multiple methods to collect data that effectively captures various manifestations of dark patterns.
    \item \textit{Complexity of Longitudinal Studies}: While longitudinal studies\cite{bolger2012power} are valuable, they pose challenges due to extended timelines, analytical complexity, and the dynamic nature of digital interfaces, which may affect the reliability and generalizability of the findings.
    \item \textit{Ethical Considerations}:With increasing regulatory focus, ethical considerations are paramount. Flexible ethical guidelines must accompany the taxonomy, requiring a dynamic approach during development.
\end{itemize}

In conclusion, our study lays a foundation for research on dark patterns. However, ongoing, multidisciplinary efforts are needed to refine  to accommodate the evolving nature of dark patterns.

\section{Related Work}
In this section, we review previous research related to dark patterns, focusing on their classification and detection. We first summarize the studies that have developed taxonomies to categorize dark patterns, as well as those that have introduced techniques to detect these patterns in graphical user interfaces (GUIs). The following subsections will delve into classification and detection in detail.
\subsection{Classification of Dark Patterns}

Numerous empirical studies have explored the concept of dark patterns across various digital platforms, including websites, mobile applications\cite{di2020ui,gray2018dark,Mathur2019}, and domains such as gaming\cite{Aagaard2022}, social media platforms\cite{habib2022identifying}, e-commerce websites\cite{Mathur2019}, cookie banners\cite{gray2021dark,nguyen2022freely}, and advertising\cite{gak2022distressing}. These empirical studies typically collect and analyze user interfaces from desktop or mobile applications. Researchers then integrate their findings to identify and categorize the types of dark patterns embedded in these interfaces.

The journey began in 2010, when Harry Brignull\cite{brignull2010types} established the concept of dark patterns, defining them as manipulative techniques used by websites and applications to push users into unintended actions, such as making purchases or signing up for services. Brignull also created a Twitter account\cite{brignull2010} where users could report and discuss instances of dark patterns they encountered in their daily digital interactions. Subsequently, Gray et al.\cite{gray2018dark} conducted manual data collection over two months on popular online platforms. They extended Brignull's concept by categorizing dark patterns into six strategic categories, such as nagging, obstructing, and sneaking. Building on these taxonomies, Mathur et al.\cite{Mathur2019}conducted a large-scale empirical study focused on approximately 11,000 shopping websites. Their approach involved simulating user shopping behavior and clustering collected website segments. Unlike Mathur et al., who focused on e-commerce platforms, Di Geronimo et al.\cite{di2020ui} examined the presence of dark patterns in 240 widely used Android applications. They recorded ten minutes of usage for each application and adopted the taxonomy developed by Gray et al.\cite{gray2023towards}, identifying 16 new types of dark patterns encompassing 31 cases.

The study most closely related to ours is by Gray et al.\cite{gray2023towards,Gray2024}, who synthesized a taxonomy of 64 dark pattern types across five categories by integrating ten existing dark pattern taxonomies. Our taxonomy, however, includes a broader range of types, with six major categories comprising a total of 106 dark pattern types. Additionally, we have innovatively annotated each type of dark pattern in our taxonomy with its harm level, application scenarios, and specific examples, enabling researchers and developers to better understand the potential impacts of dark patterns across different contexts. Moreover, based on our taxonomy, we conducted an in-depth analysis of the capabilities and limitations of known dark pattern detection tools, and we provided a comprehensive assessment of the quality of existing datasets, thus identifying gaps in current detection and remediation approaches. This provides clear directions for future research and tool development.

\subsection{ Detection of Dark Patterns}

Dark pattern detection techniques are evolving rapidly, progressing from initial manual explorations to semi-automated clustering methods, text-based classification techniques, traditional machine learning approaches, and more recently, deep learning methods that integrate computer vision. Additionally, methods leveraging large language models’ contextual understanding for detection are beginning to emerge.

Dark pattern detection initially relied on manual exploration, primarily conducted by domain experts\cite{brignull2010types,di2020ui,gray2018dark,habib2022identifying}. To address the limitations of manual exploration, Mathur et al.\cite{Mathur2019}introduced semi-automated techniques that simulate user behavior to identify dark patterns on shopping websites. They employed clustering to group related UI patterns, enhancing detection efficiency. Nazarov et al.\cite{nazarov2022clustering} compared two clustering methods, analyzed 12 dark patterns, and provided feature descriptions for each category, laying the groundwork for future dark pattern site analysis services. Unlike clustering methods, traditional machine learning approaches further automated detection. For instance, Curley et al.\cite{curley2021design} combined natural language processing (NLP) and web-crawling technologies to develop a web-based automated dark pattern detection framework. Subsequently, deep learning methods incorporating detection capabilities began to emerge. Yada et al.\cite{Yada2022DarkPI} used Mathur's\cite{Mathur2019}dataset to compare traditional machine learning methods with a transformer-based model they proposed for dark pattern detection, achieving an accuracy of 0.975. Mansur et al.\cite{mansur2023aidui} employed computer vision and NLP techniques to identify visual and textual cues in app screenshots, successfully detecting ten types of dark patterns, and developed the ContextDP dataset for evaluation, achieving an overall precision of 0.66, recall of 0.67, and F1 score of 0.65. Chen et al.\cite{chen2023unveiling} proposed a knowledge-driven system, UIGuard, combining computer vision and NLP technologies that performed well in dark pattern detection, achieving a precision of 0.82, recall of 0.77, and F1 score of 0.79. Kodandaram et al.\cite{Kodandaram2023} used a multimodal model to detect disguised ads on webpages, with tested F1 scores of 0.86 and 0.88. Recently, as large language models have developed, Sazid et al.\cite{Sazid2023}utilized their contextual understanding to detect some dark patterns based on Mathur's\cite{Mathur2019} and Yada's\cite{Yada2022DarkPI} datasets, achieving favorable performance.

Building on prior work, our study evaluates the capabilities and limitations of existing tools for detecting dark patterns as defined in our taxonomy. We also examine various benchmark datasets used for dark pattern detection.

\section{CONCLUSION}

In conclusion, this research represents a significant milestone in the rapidly evolving field of dark pattern research. Through a meticulous systematic review, we identified 113 seminal papers that enabled us to construct the most comprehensive standardized taxonomy of dark patterns to date. This taxonomy not only includes the types and descriptions of each dark pattern and their categorization but also annotations on their impact on users, potential application scenarios, and specific examples. Additionally, we leveraged the taxonomy to evaluate the capabilities of existing dark pattern detection tools and the availability of current data, and we constructed a standardized dataset. Our taxonomy serves as both a standard terminology for categorizing dark patterns and as a practical tool for researchers, practitioners, and policymakers.

The road ahead is challenging, requiring multidisciplinary approaches and continual refinement of methods and tools. The field of dark patterns is constantly evolving, influenced by rapid technological advancements and shifts in social norms. Thus, future research must focus on developing more robust methods, including open-source tools and ethical guidelines, to adapt to this ever-changing landscape.
\section*{Acknowledgments}
We thank the reviewers for their insightful comments and suggestions. This work was partially funded by the National Natural Science Foundation of China (61972359, 62132014), Zhejiang Provincial Key Research and Development Program of China (2022C01045), Zhejiang Provincial Natural Science Foundation of China (LQ23F020020), the Education Department of Hunan Province (21B0313), the Natural Science Foundation of SZTU Top Talent (GDRC202132), the SZTUEnterprise Cooperation Project (20221061030002, 20221064010094), the Inner Mongolia Key Project (2021ZD0044), and the SZTU Project (20224027010006).

\bibliographystyle{ACM-Reference-Format}
\bibliography{References}




\end{document}